\documentclass[useAMS,usenatbib,fleqn]{mn2e}

\usepackage{psfig}
\usepackage{epsfig}
\usepackage{lscape}
\usepackage{rotating}
\usepackage{amsmath}
\usepackage{graphicx,color}
\usepackage{subfigure}
\usepackage{multicol}
\usepackage{multirow}
\usepackage{amssymb}
\usepackage{comment}
\usepackage{textcomp}
\usepackage{pslatex}
\usepackage{graphicx,color,layout}
\usepackage{multirow}

\newcommand{\psra}{PSR J1738$-$2330}
\newcommand{\psrb}{PSR J1752+2359}
\newcommand{\pa}{J1738$-$2330}
\newcommand{\pb}{J1752+2359}

\title[On the long nulls of PSRs J1738$-$2330 and J1752+2359]{On the long nulls
of PSRs J1738$-$2330 and J1752+2359}

\author[Vishal Gajjar et al.]
{Vishal Gajjar$^{1}$\thanks{e-mail:gajjar@ncra.tifr.res.in}, 
B. C. Joshi$^{1}$, Geoffrey Wright$^{2}$\\\\
$^{1}$National Centre for Radio Astrophysics, Post Bag 3, Ganeshkhind, Pune 411
007, India\\
$^{2}$Astronomy Centre, University of Sussex, Falmer, Brighton BN1 9QJ, UK.}

\date{\today}

\pagerange{\pageref{firstpage}--\pageref{lastpage}} \pubyear{2009}

\begin{document}

\maketitle

\label{firstpage}

\begin{abstract}
This paper compares and contrasts the emission of two high nulling fraction
pulsars, PSRs J1738$-$2330 and J1752+2359. In both pulsars the emission bursts appear in a
quasi-periodic fashion with typical separations of several hundred pulses, and
in \pa\ there is evidence of two underlying periodicities with memory
persisting for at least 11 bursts. By contrast, in \pb\ the pattern coherence
is rapidly lost and the burst/null  lengths appear to be selected randomly from
their respective quasi-normal distributions. 
The typical emission bursts of \pa\ exhibit a steady 
exponential decay of on-pulse energy accompanied by a flickering emission
characterized by short frequent nulls towards their end. In the bursts of \pb\
the flickering is absent, the decay more pronounced and the energy released
during each bright phase is approximately constant. Unlike \pa, the average
profiles for the first and the last pulses of \pb\ bursts differ slightly from the
pulsar's overall profile, hinting at differences between the two pulsars in
their transitions from null to burst state (and vice-versa). During its long
null phases, \pb\ is found to emit random weak inter-burst pulses (IBPs)
whose profile peak is somewhat offset with respect to the overall average
profile. Such pulses have no equivalent in \pa\ or in any known pulsar
hitherto. They may pervade the entire emission of this pulsar and have a
separate physical origin to normal pulses. On the basis of our comparison we
conclude that a pulsar's nulling fraction, even when high, remains a poor
guide to its detailed subpulse behaviour, as previously 
found for pulsars with small nulling fractions.
\cite[]{gjk12}. 
\end{abstract}


\section{Introduction}
\label{intro}
Pulse nulling, the cessation of pulse emission 
from a single pulse to several thousands of pulses and which is found in about 
100 pulsars, has defied satisfactory explanation since its 
discovery \cite[]{bac70} more than four decades ago. 
Nulling pulsars exhibit a variety of nulling fractions 
(NF)\footnote{the fraction of pulses for which 
no detectable radio emission is seen} and one might hope that this apparently 
fundamental parameter could be used to characterise further aspects of the 
pulsar's behaviour. However, a recent study dashed this hope and 
concluded that pulsars with similar low NFs ($\sim$ 1 \%) are not necessarily
similar in their nulling pattern\footnote{the arrangements 
of null and burst pulses in a single pulse sequence}
\cite[]{gjk12}. In this paper we set out to see if 
this result still holds even for pulsars with large 
NF ($>$ 80 \%). We compare and contrast two high-NF 
pulsars and assess to what extent their nulling 
patterns follow a common statistical rule.

The first of our pulsars, \psrb, was discovered in a high Galactic
latitude survey with the Arecibo telescope \cite[]{fcwa95}. Subsequent 
timing observations indicated
interesting single pulse behaviour, with
bursts of up to 100 pulses separated by 
nulls of about 500 pulses \cite[]{lwf+04}, giving an NF of about 80 \%. 
This study also noted an intriguing exponential decrease in the pulse 
energy during a burst, a feature previously seen in only 
a few nulling pulsars \cite[]{rw08,bgg10,lem+12}. 
The second, \psra, was discovered in 
the Parkes multi-beam pulsar survey \cite[]{lfl+06}. 
A study of this pulsar at 325-MHz gave a lower limit 
of 69 \% to its  NF \cite[]{gjk12}, and -- not unlike 
\pb\ -- an intermittent pattern of quasi-periodic 
bursts interspersed with long nulls was reported. 
Table \ref{paratable} compares the basic parameters 
for the two studied pulsars. Note that \pa\ is 
approximately five times slower and three times 
younger than \pb. 

In previous studies of these pulsars the typical duration for single pulse
observations was 1-2 hours. Such short observations, particularly 
for pulsars with periods around 1 s, do not yield 
sufficient nulls and bursts for a satisfactory comparison 
of their statistical properties. In this study 
much longer observations were undertaken to obtain 
a large sample of nulls and bursts in each pulsar. 

\begin{table*}
{\small
 \caption{Basic parameters$^{a}$ and observational details for the two
  observed pulsars. Columns give the pulsar name at 2000 epoch, period (P), dispersion measure
  (DM), characteristic  age ($\times$10$^6$ Yr), surface magnetic field (B$_s$), date
  of the observations, telescope, sampling resolution (degrees/sample) and length of the observations
  (pulses).}
 \label{paratable}
 \centering 
 \begin{center}
 \begin{tabular}[ht]{|l|c|c|c|c|c|c|c|c|c|}
 \hline 
 \multirow{3}{*}{PSRs} &    &                    &           &        &       &     &           &            &   \\
 \cline{2-10} \\
              &    P    &  DM 		           & Age       & B$_{s}$ & $~~~~~$  & Date & Telescope & Resolution       & Length \\
              &  (sec)  &  (pc$\cdot$cm$^{-3}$)   & (MYr) & ($\times$10$^{12}$G) & &      &           & (degrees/sample) & (Pulses) \\              
 \hline       
 J1738$-$2330 &  1.98   & 99.3                    & 3.6       & 4.16	 & $~~~~~$     & 2010 Oct 24   & GMRT  &  0.18  & 8463   \\
 J1752+2359   &  0.41   & 36.0                    & 10.1      & 0.52  	 & $~~~~~$     & 2010 Sep 3  & GMRT  &  0.87  & 67601  \\ 
              &         &                         &           &        & $~~~~~$     & 2006 Feb 12  & Arecibo & 0.36 & 4891   \\
 \hline
 \end{tabular}
 \end{center} 
 \hfill{}
 \begin{picture}(0,0)
  \put(-460,0){$^{a}$ ATNF Catalogue : www.atnf.csiro.au \cite[]{mhth05}}
  \put(-370,83.5){Basic parameters$^a$}
  \put(-150,83.5){Observations}
 \end{picture}
 }
\end{table*}
The observations and analyses are described in Section \ref{observations}.  
The nulling patterns and their quasi-periodicities are discussed  
in Sections \ref{qperiodsection} and \ref{pcf}. The variations in burst pulse
energy and the modelling thereof is described in Section \ref{BBB_patten}. 
The emission behaviour during null-to-burst transitions and vice-versa are 
discussed in Section \ref{first_and_last_bbb_section}. Unusual emission 
behaviour present in the null sequences of \psrb\ is analysed in  
Section \ref{emission_in_null}. Section \ref{Arecibo} discusses 
the polarization profiles of \psrb. The results of this study and their 
implications are summarized in Sections \ref{conclusion} and 
\ref{discussion}, respectively. 
In addition, Appendix \ref{gps} examines the possible presence of  giant pulses
in \psrb. Supporting material on the pair correlation function and 
modelling used in the paper is provided in Appendices \ref{apppcf} 
and \ref{appa}.  

\section{Observations and analysis}
\label{observations}
\begin{figure}
  \centering
  \subfigure[]{ 
  \includegraphics[width=1.5in,height=4in,angle=0,bb=0 0 193 499]{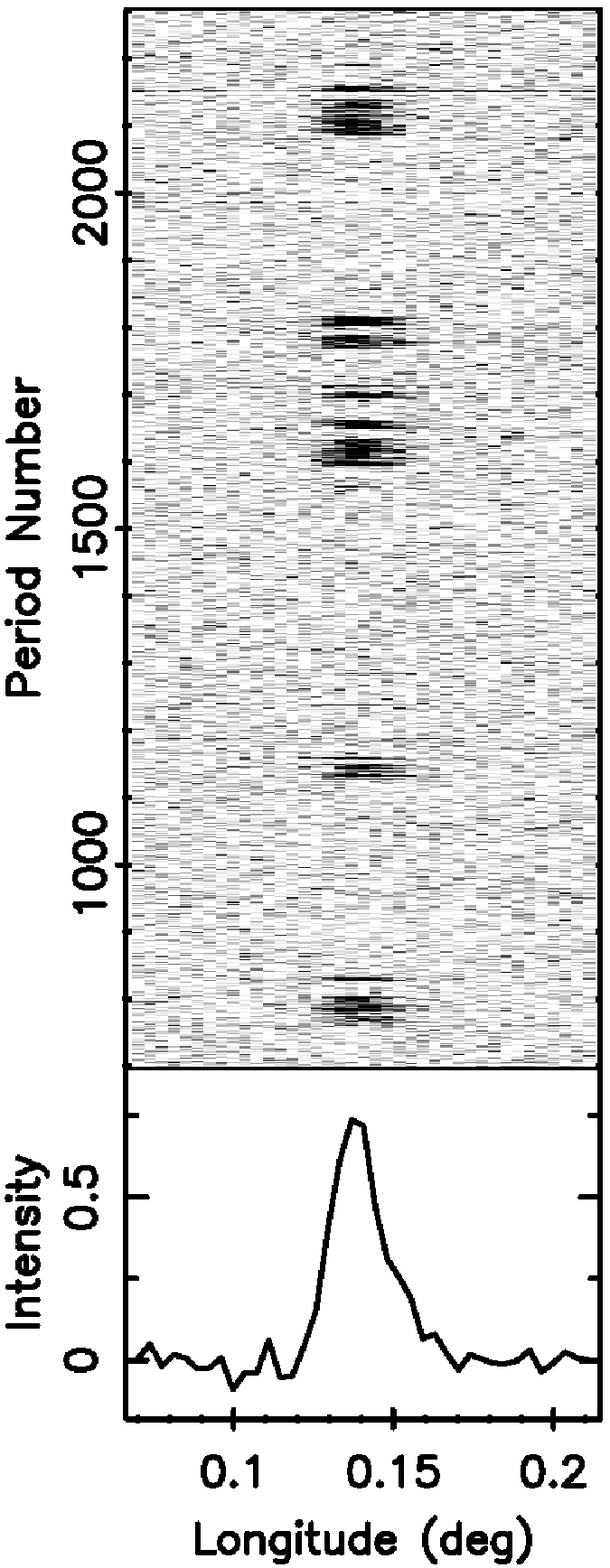}
  \label{j1738spdisplay}
  }
  \subfigure[]{
  \includegraphics[width=1.5in,height=4in,angle=0,bb=0 0 193 499]{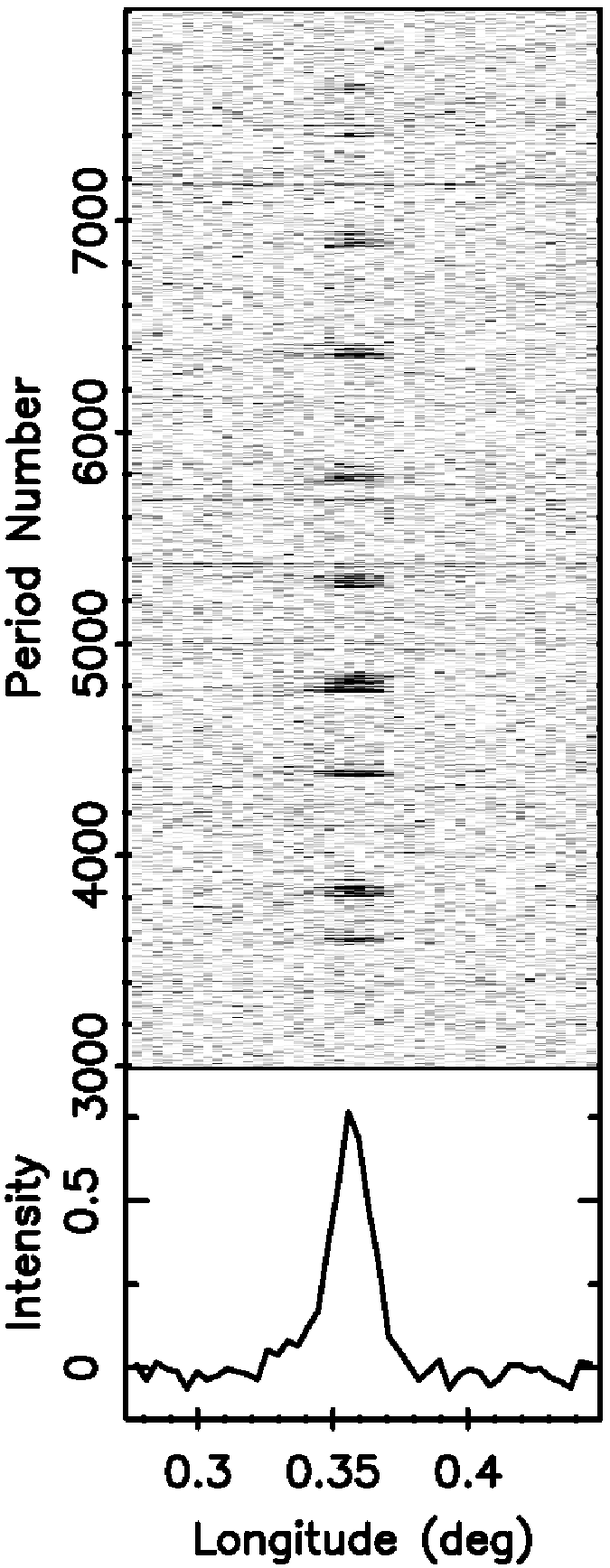}
  \label{j1752spdisplay}
  }
  \begin{picture}(0,0)
  \centering 
  \put(-190,290){PSR J1738$-$2330}
  \put(-75,290){PSR J1752+2359}
  \end{picture}
 \caption{Single pulse sequences in grey-scale intensities for 
 both pulsars observed at 325-MHz with the GMRT. 
 (a) A sequence of around 1500 consecutive 
 pulses from \pa. Flickering short nulls towards the end of 
 bright phases can be seen for this pulsar.  
 (b) A consecutive sequence of around 5000 pulses 
 from \pb. The quasi-periodic pattern 
 of bright phases is clearly evident.}
 \label{spdisplay}
\end{figure}

\begin{figure*}
 \centering
 \subfigure[]{
 \includegraphics[width=2.5 in,height=3.1 in,angle=-90,bb=30 25 588 768]{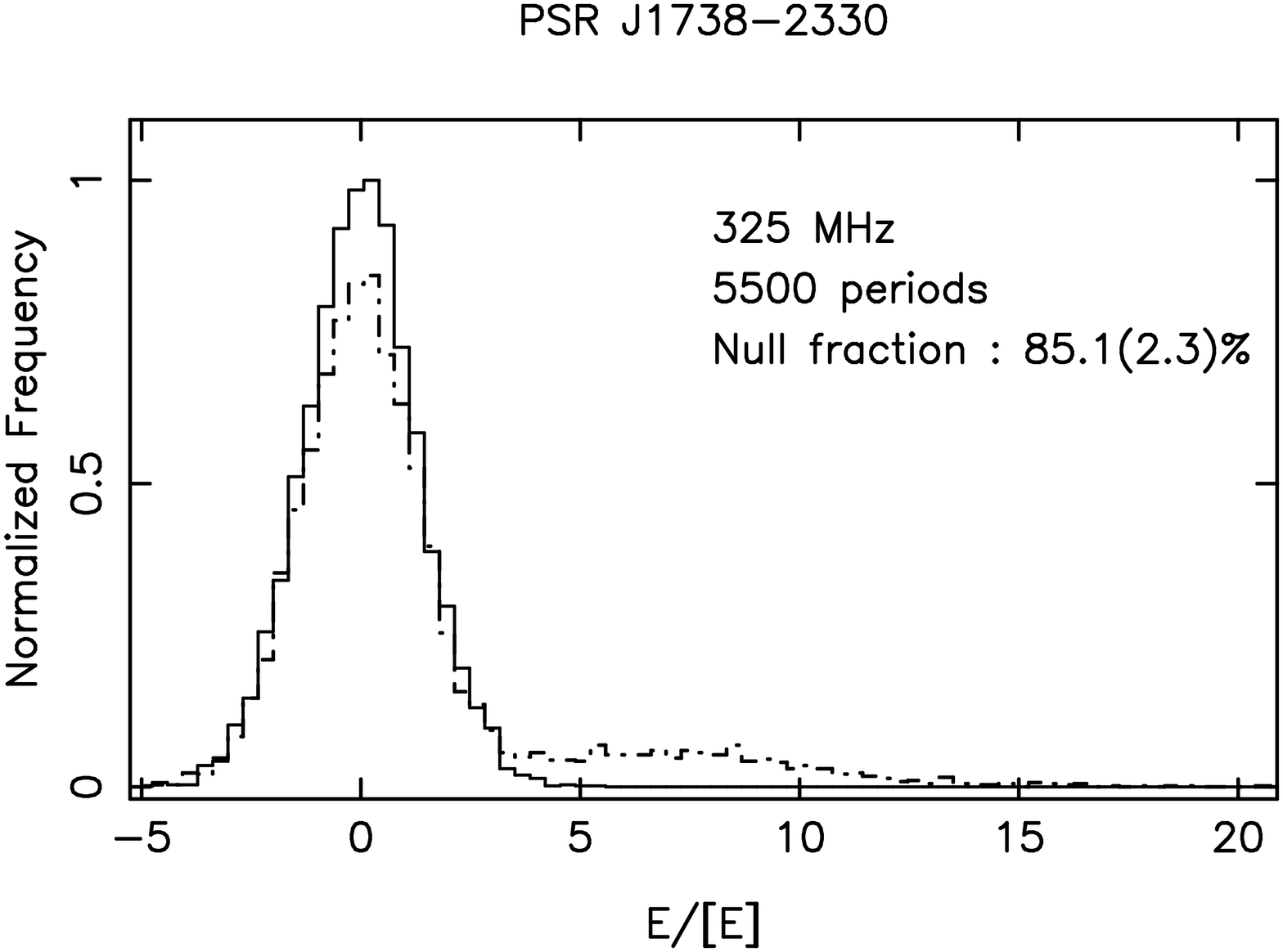}
 \label{j1738_hist}
 }
 \centering
 \subfigure[]{
 \includegraphics[width=2.5 in,height=3.1 in,angle=-90,bb=30 25 588 768]{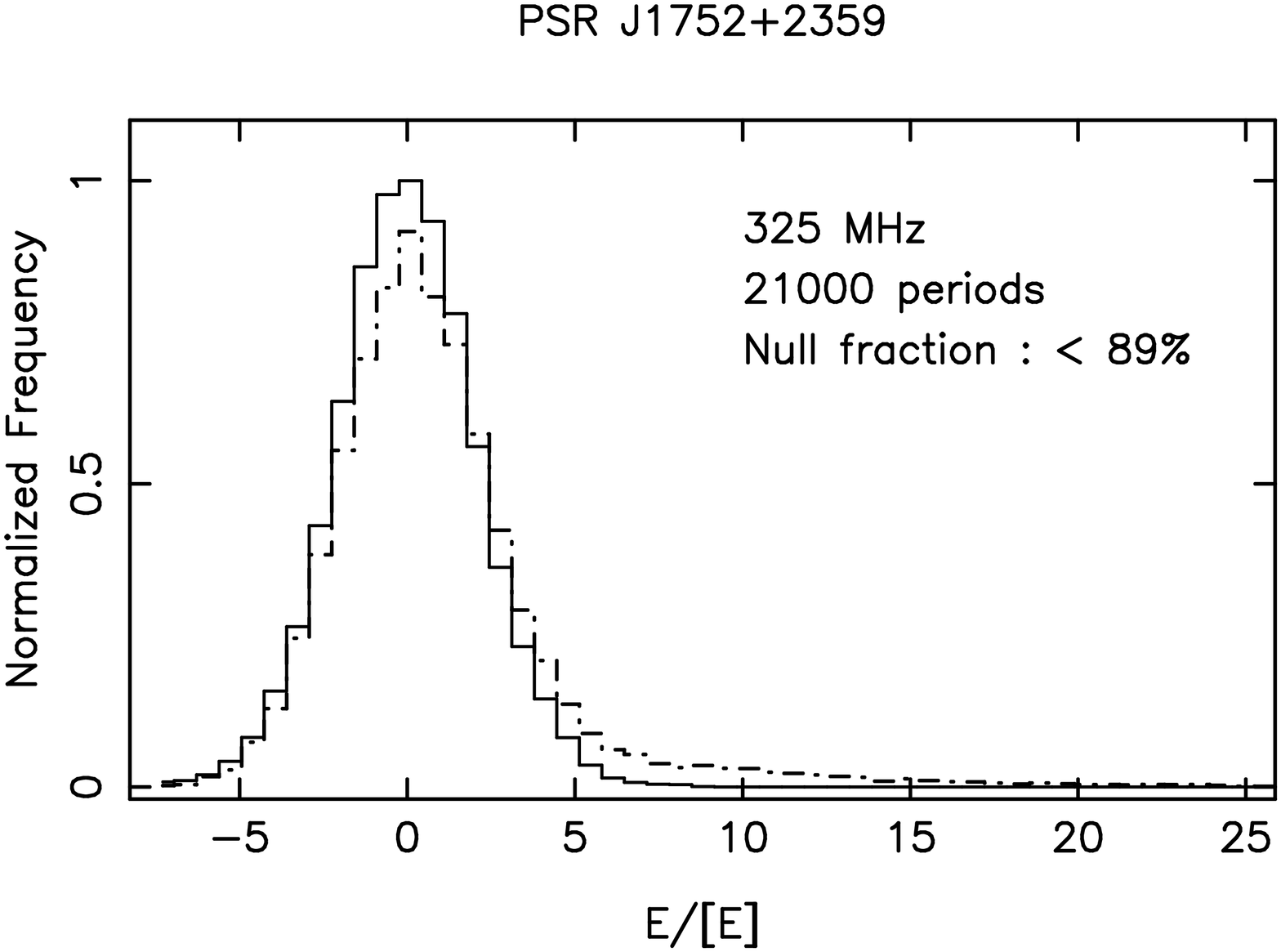}
 \label{j1752_hist}
 }
  \caption{On-pulse (dash-dot-dash line) and off-pulse energy 
 (solid line) histograms for PSRs (a) J1738$-$2330 and 
 (b) J1752+2359 obtained with the GMRT observations at 325-MHz. 
 Both pulsars show the large fraction of null pulses. \psra\ shows
 clear bimodal burst pulse distribution compared to 
 the smoother on-pulse intensity distribution seen in \pb.
 The NFs were estimated using the method discussed in \protect\cite{rit76} and
 \protect\cite{gjk12}. For \pb\ we can only estimate an upper limit on the NF of around $<$89\%  
 as some fraction of weak burst pulses will be included in the  null pulse
 distribution. 
}
\label{hist_both_fig}
\end{figure*}

The observations of both pulsars were carried 
out with the Giant Meterwave Radio Telescope (GMRT)
at 325-MHz. We also analysed a small section 
of observations (only for \pb) observed at 327-MHz with  
the 305 m Arecibo Telescope in Puerto Rico 
(Rankin, private communication). 
The details regarding the date, 
the sampling resolution  and the total number of observed 
pulses are listed in Table \ref{paratable}.

The GMRT consists of 30 dish antennas, each 45 m 
in diameter \cite[]{sak+91}. We used the GMRT 
in a single beam phased array mode by only including the 
short spacing ($\sim$ 4 km) antennas to minimize  
the fluctuations in their phase differences due 
to the ionosphere. All central square antennas (14 
antennas) in addition to the first two antennas from 
each arm (6 antennas) were used. Few antennas needed 
to be dropped from the configuration due to bad bandshapes and/or 
radio frequency interference (RFI). Thus, signals from 
typically 16 to 19 short spacing antennas were added 
in phase during the observations of both pulsars. 
Both pulsars were observed for around 8 hours. 
Radio waves were received by the front-end system and 
then transferred to the central electronics building 
using an optical fibre link. At the central electronics 
building, signals were processed by the GMRT software 
backend, which is a real-time correlator for 32 
antennas$\times$2 polarizations with 33.3-MHz 
bandwidth \cite[]{rgp+00}. Voltages from each antenna 
were added after phase compensation to form a coherent 
sum, which was squared to obtain the total intensities. 
The data were recorded in this mode with an effective 
integration time of around 1 ms for both  pulsars. 
Phase equalisation was carried out every 1.5 to 2 hours 
during the observations due to short instrumental 
phase stability time scales at lower frequencies. Hence, 
8 hour long observations for each pulsar 
were broken up into 4 to 5 different sessions and data 
were recorded separately for each session. 
Data were then converted to SIGPROC\footnote{http://sigproc.sourceforge.net}
filterbank format for off-line processing. 

During the off-line analysis 
a few channels briefly showed the presence of narrow-band RFI.  
To keep a similar sensitivity level during each individual session, we 
flagged all the RFI affected channels from the entire 
observing session.  Remaining channels were dedispersed 
using the respective nominal DM (given in Table \ref{paratable}) 
for both pulsars. The dedispersed time series were folded to 256 longitude 
bins for every single period for both pulsars using the rotation period obtained from the
polycos\footnote{http://www.atnf.csiro.au/people/pulsar/tempo/}. 
These single pulse datasets were used for rest of the analysis. 
Small sections of the observed single pulses in 
grey-scale intensities are shown in Fig. \ref{spdisplay} 
for both pulsars. The on-pulse and the off-pulse 
energy histograms are shown in Fig. \ref{hist_both_fig}. 
Note that no giant pulses are seen for \pb\  
in our observations at 325-MHz, contrary to reported giant pulses with energies 
of about 200 times the mean pulse energy at 111-MHz \cite[]{EK05}. 
A discussion of this result can be found in Appendix \ref{gps}. 

For \psrb\ around 30 minutes of archival full polar 
observations at 327-MHz from the Arecibo
telescope was also analysed to confirm the presence of weak 
burst pulses during the null states and to compare 
linear and circular polarization profiles. 
Details regarding the data reduction 
and the polarization calibration are similar 
to those discussed by \cite{rwb13}. 

\section{Null-burst statistics} 
\label{quasiperiod}
\label{qperiodsection}
 \begin{figure}{
 \centering
 \subfigure{
  \includegraphics[width=2.5 in,height=2 in,angle=0,bb=100 100 360 302]{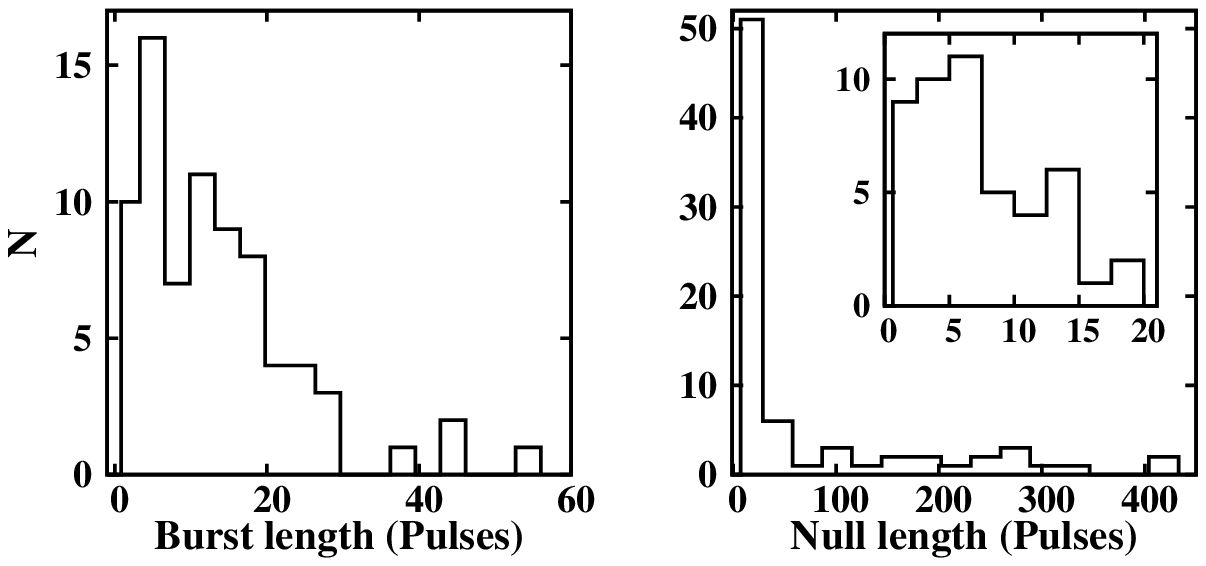}
 \label{nbhist_j1738}
 }
 \vspace{0.5 cm}
 \begin{picture}(0,0)
  \centering 
  \put(-120,115){\large PSR J1738$-$2330}
  \put(-120,-27){\large PSR J1752+2359}
  \put(-150,-15){(a)}
  \put(-30,-15){(b)}
  \put(-150,-160){(c)}
  \put(-30,-160){(d)}
 \end{picture}
 \subfigure{
 \includegraphics[width=2.5 in,height=2 in,angle=0,bb=100 70 360 272]{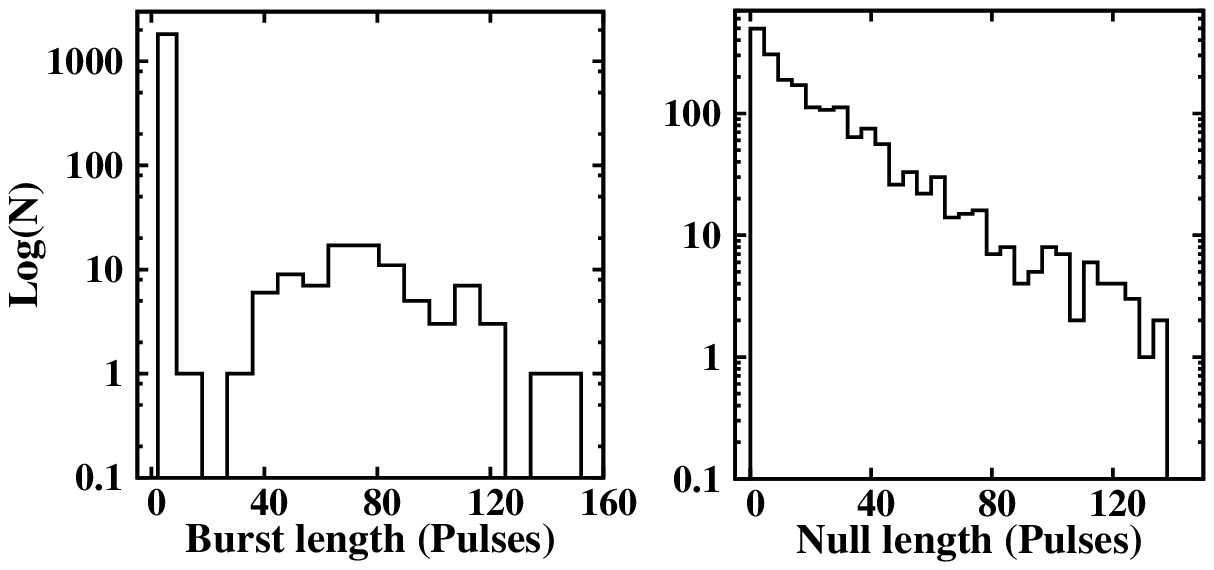}
 \label{nbhist_j1752}
 }
\caption{Plot of conventional burst and null length histograms (BLH and NLH) 
 for PSRs \pa\  and \pb\ obtained from the GMRT observations. The NLH (b) 
 for \pa\ is shown with an inset plot depicting the distribution of 
 short nulls which is very similar to the distribution of the bursts in 
 (a), indicating that the high NF of this pulsar arises from an excess 
 of long nulls. The NLH (d) and the BLH (c) for \pb\ are shown with 
 the measured counts on a log-scale to bring out the details for longer 
 bursts and nulls. Note the large number of single bursts in (c), whose 
 random occurrence among the long nulls generates an exponential distribution in
 the NLH (d). Thus the two pulsars have very different null distributions despite
 similar NFs.}
\label{nbhist_both}
}
\end{figure}

The single pulse sequences of Fig. \ref{spdisplay} show the burst pulses of both
pulsars clustering together in groups, which we will refer to as the {\it bright phases}
interspersed with long null phases (the {\it inter-burst} or {\it off-phases}),
giving a quasi-periodic effect. 

However, behind this general similarity we find a number of important
differences between the two pulsars, and these can be seen in the burst and null
length histograms of each pulsar (Fig. \ref{nbhist_both}). In \pa, the
lengths of the bursts (see Fig. \ref{nbhist_both}a) are much shorter than those typical of \pb\
(shown in Fig. \ref{nbhist_both}c) because in its bright phases the bursts of the former are equally
mixed with short nulls, as can be seen from the inset of short nulls in Fig. \ref{nbhist_both}b.
Together this results in bright phases of up to 60 pulses in length, with the
nulls predominating towards the end (see Section \ref{BBB_patten}). The high NF
of this pulsar then arises through the presence of a long tail of long nulls in
the histogram of Fig. \ref{nbhist_both}b. All the designated null pulses of this pulsar, whether
occurring within or between bright phases, were integrated and found to show no
profile of significance. 

By contrast, in \pb\ the bright phases consist of sustained non-null pulses
(see Fig. \ref{nbhist_both}c), typically of 70-80 pulses. What is unusual about this pulsar is the
very large number of isolated burst pulses which occur in the inter-burst
phases. These are not evident in Fig. \ref{spdisplay} and can only be found by a careful
inspection of the sequences. We designate them as inter-burst pulses (IBPs) and
show in Section \ref{emission_in_null} that they appear at random during the
inter-burst phases, and maybe throughout the entire emission of \pb. As a
result, the apparently long nulls of this pulsar become subdivided in a random
way giving rise to the exponential distribution strikingly visible in the 
null length histogram (NLH;  see Fig. \ref{nbhist_both}d). No such effect is seen in \pa.

One consequence of the burst-null mix in the bright phases of \pa\  and the
IBPs in \pb\ is that the conventional burst and null length histograms of both
pulsars in Fig. \ref{nbhist_both} show no evidence of the quasi-periodic
behaviour of the bright and off-phases despite it being very clear in the pulse
sequences shown in Fig. \ref{spdisplay}. To overcome this, we carried out a visual
inspection of the single pulses of both pulsars and identified appropriate
bright phases and their separation (the separation being defined as the
number of pulses between the first pulses of two consecutive bright phases). 



\begin{figure}
 \centering
  \subfigure[]{
  \centering
  \includegraphics[width=1.5in,height=1.4in,bb=80 50 430 302]{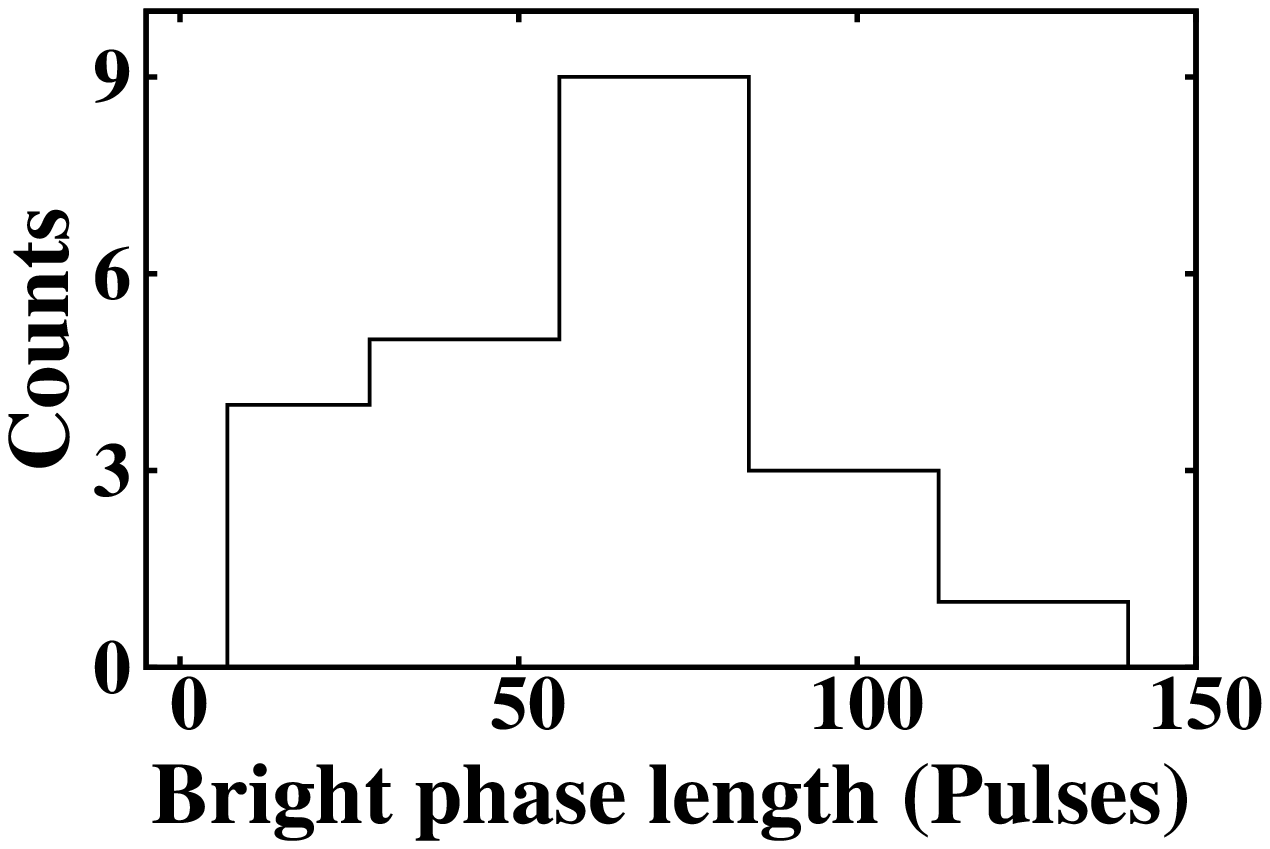}
  \label{j1738BBB_length}
  }
  \subfigure[]{
  \centering
  \includegraphics[width=1.5in,height=1.4in,bb=50 50 400 302]{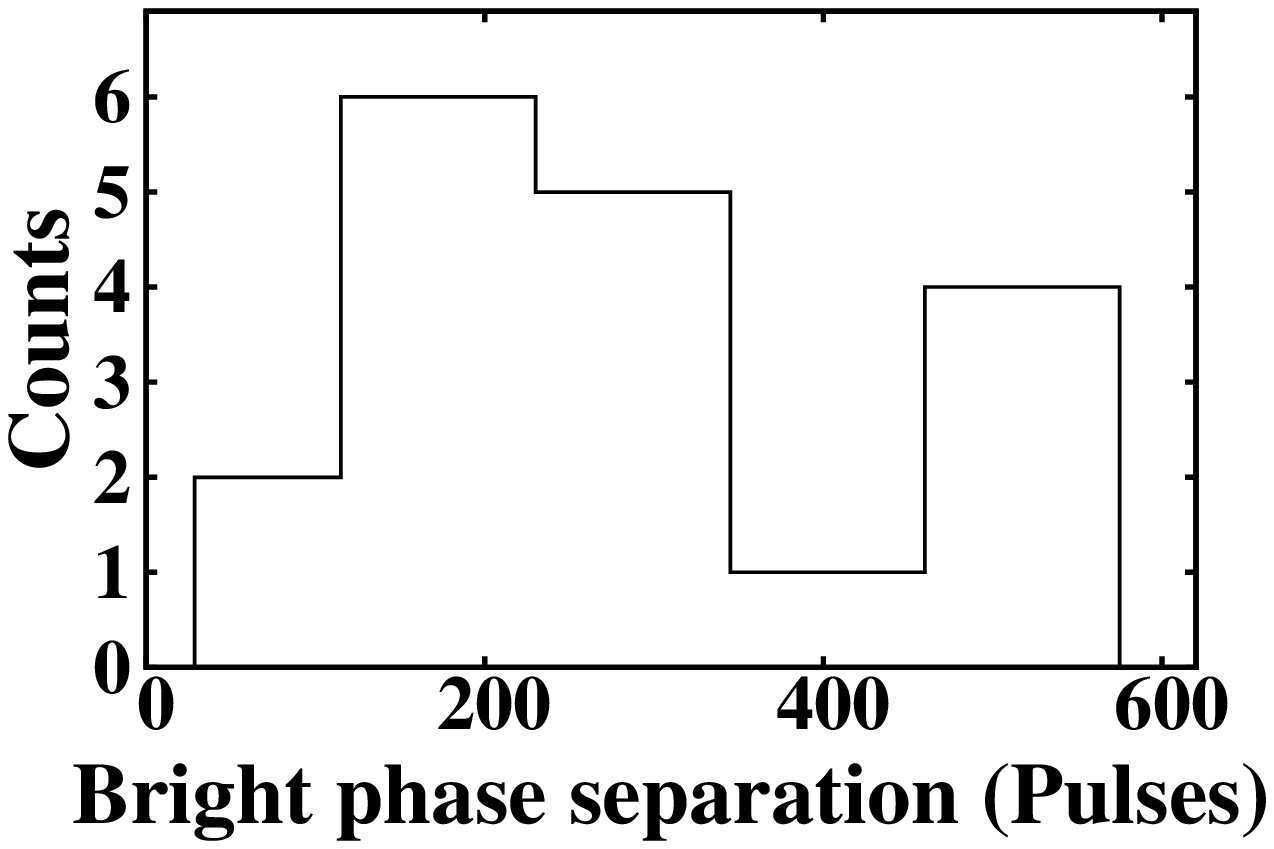}
  \label{j1738BBB_gap}
  }
  \begin{picture}(0,0)
  \centering 
  \put(-145,100){\large PSR J1738$-$2330}
  \put(-145,-25){\large PSR J1752+2359}
  \end{picture}
  \hfill
  \subfigure[]{ 
  \includegraphics[width=1.5in,height=1.4in,angle=0,bb=80 60 430 312]{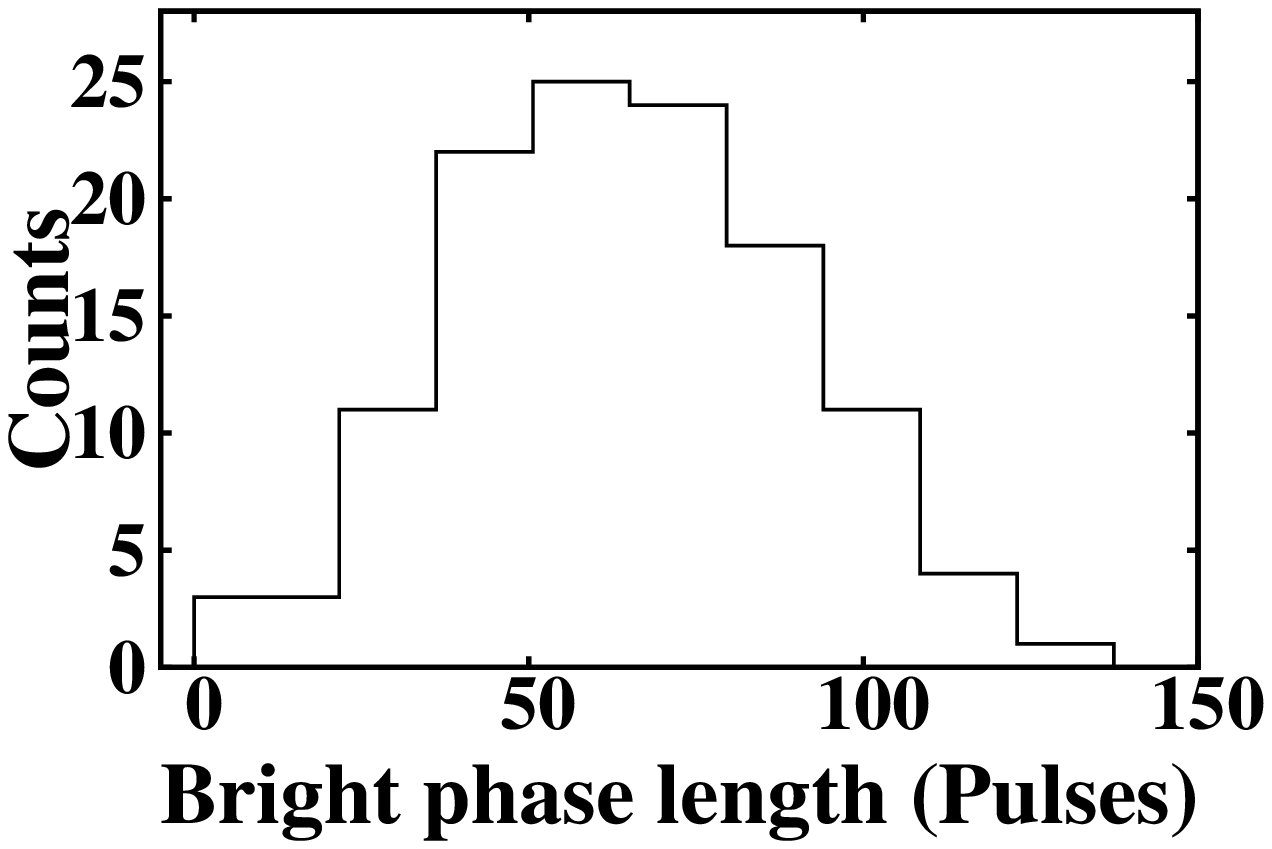}
  \label{j1752BBB_length}
  }
  \centering
  \subfigure[]{
 \includegraphics[width=1.5in,height=1.4in,angle=0,bb=50 60 400 312]{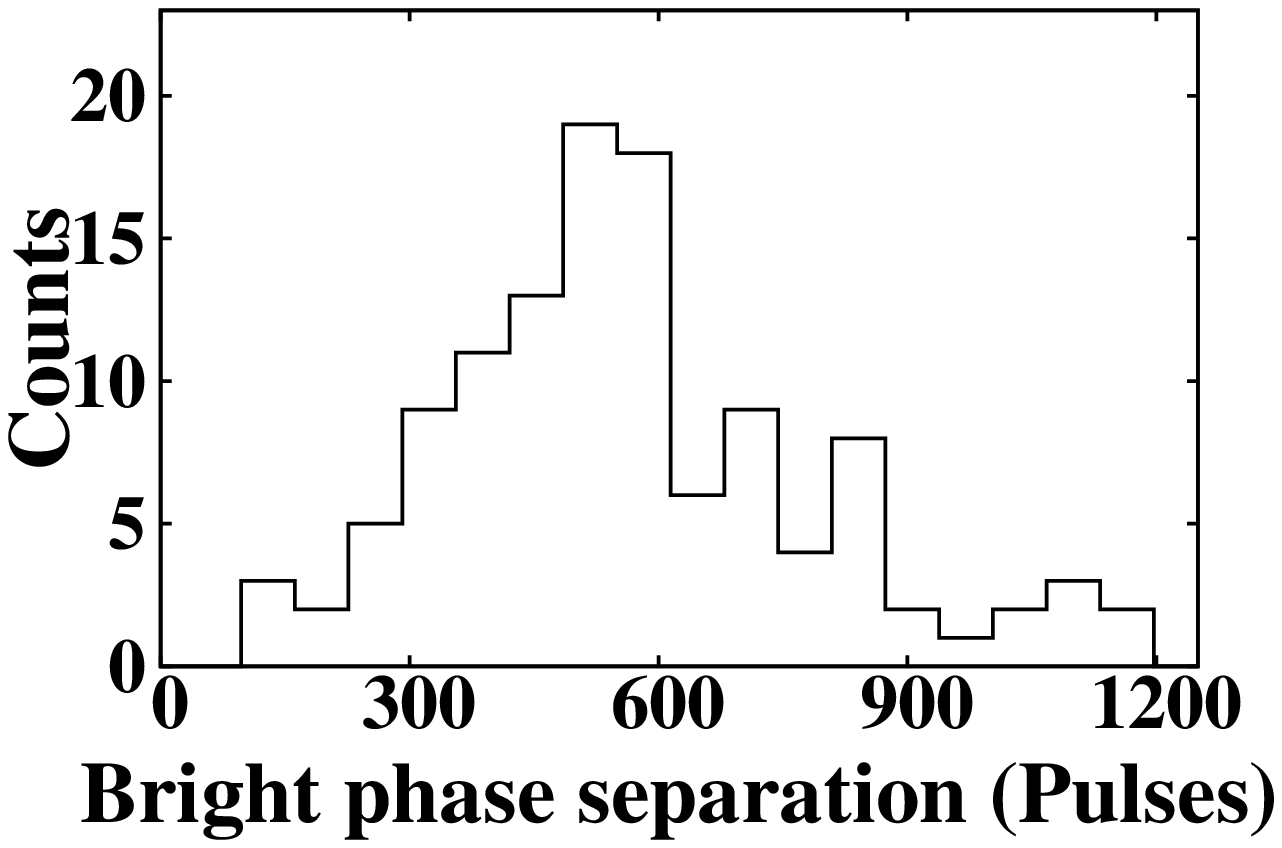}
  \label{j1752BBB_gap}
  }
 \caption{Histograms for the length of the visually identified bright phases
 for PSRs (a) \pa\ (c) \pb\ obtained from the 325-MHz observations at the GMRT. 
 The histograms of separation between the first  
 pulse of successive bright phases for PSRs (b) \pa\ (d) \pb\ are also shown.  
 The distributions for bright phase length are similar, while 
 those for bright phase separation are very different.}
 \label{BBB_lengh_gap_fig} 
 \end{figure}

In the case of \pa, 21 bright phases were identified and a histogram of these
is shown in Fig. \ref{BBB_lengh_gap_fig}a. This distribution shows a peak at around
50 to 70 pulsar periods with a spread of around 40 pulsar periods, a result
which might be expected from combining the short bursts and nulls of Figs.
\ref{nbhist_both}a and \ref{nbhist_both}b. Likewise, a histogram of the 
separations between the first pulse of two successive bright phases is 
shown in Fig. \ref{BBB_lengh_gap_fig}b. This has a surprising bimodal character with peaks around
170 and 500 pulses. Lengths of around 500 pulses cannot be formed by combining a
typical bright phase length from Fig. \ref{BBB_lengh_gap_fig}a and the longest null
length from Fig. \ref{nbhist_both}b,
where the maximum is 400 pulses. We can therefore deduce that the longest
inter-burst phases must be interrupted at some point by very short (and maybe
weak) bursts which were rejected as burst pulses. This explains the longer
off-phase stretches seen in this pulsar's sequence of Fig. \ref{spdisplay} and
foreshadows our discussion of this pulsar's quasi-periodic patterns in Section
\ref{pcf}.

For \pb, the identification of bright phases was a more difficult task, since
it had to take into account the intrusion of inter-burst pulses (IBPs) in the
off-pulse phases and the fact that the precise closure of a fading bright phase
was sometimes difficult to fix (see Section \ref{BBB_patten}). We were able to
identify around 123 bright phases from a visual check of the single pulse
sequences in the 8-hour observations. The bright phase lengths of Fig.
\ref{BBB_lengh_gap_fig}c show much the same distribution as the burst lengths of
Fig. \ref{nbhist_both}c (apart from the single-pulse bursts). 
We find a prominent peak at around 60 pulsar periods with 
a spread of around 40 pulsar periods. The distributions of 
bright phase lengths of \pa\ and \pb\ are similar, but note that the former has a smaller 
number of bright phases. The bright phase separations of \pb\ also formed a
broad distribution around a central peak. We measured 120 examples and the
histogram is shown in Fig. \ref{BBB_lengh_gap_fig}d. The peak is
at about 570 pulses with separations ranging from 150 to 
1200 pulsar periods. The wide range of separations indicates that the nulling 
pattern is not strictly periodic, as can be seen in Fig. \ref{spdisplay}, and is
discussed in the next Section.

\begin{figure*}
  \centering
 \subfigure[]{
 \includegraphics[width=3 in,height=2.5 in,angle=0]{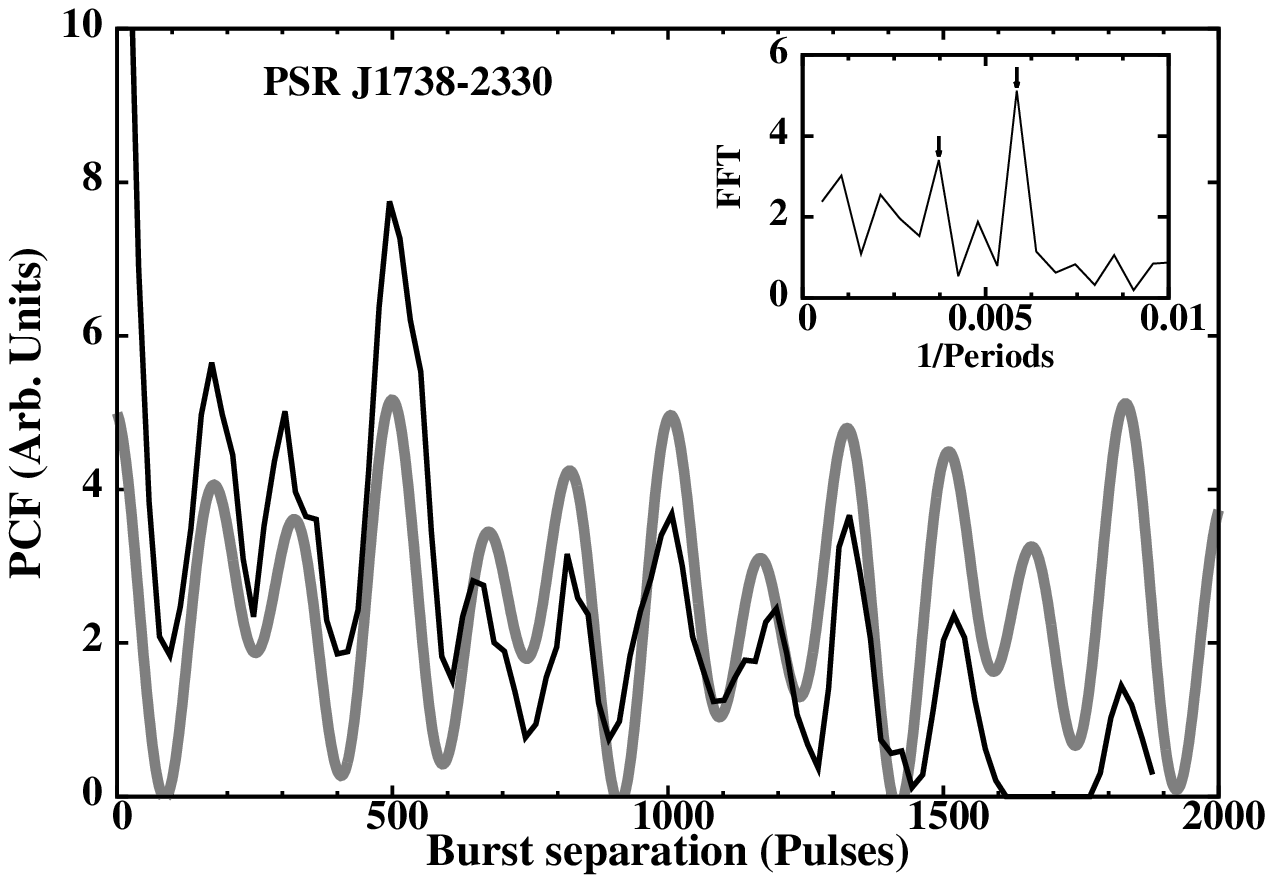}
 }
 \centering
 \subfigure[]{
 \includegraphics[width=3 in,height=2.5 in,angle=0]{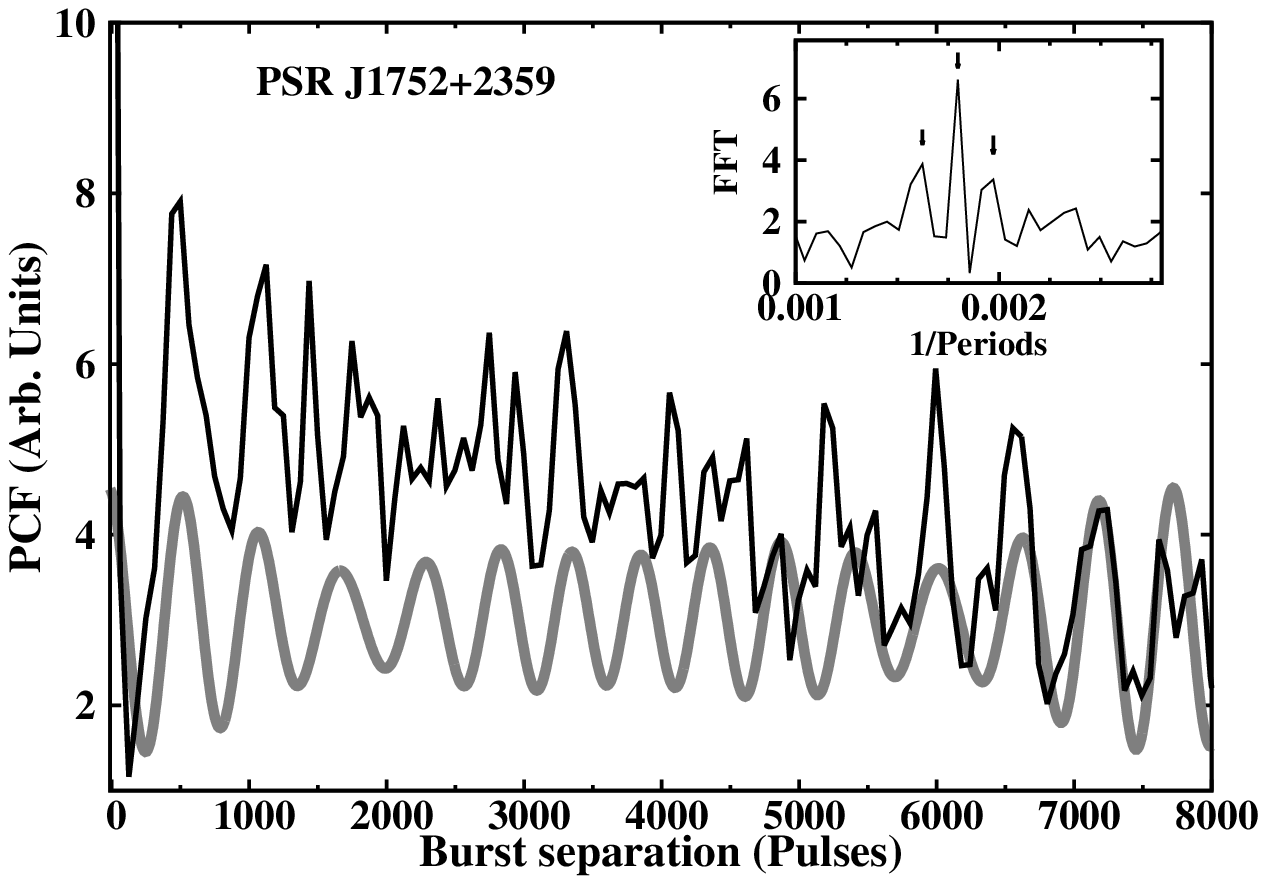}
 }
 \caption{Pair Correlation Functions for (a) \pa\ and 
 (b) \pb. The PCF is a histogram of the measured separations between observed
 burst pulses in units of pulsar periods. The reduction in amplitude of the peaks at
 large separations is due to the finite length of the observations, which in
 both cases is about one hour. The inset figures show the Fourier spectra of the
 corresponding PCF fluctuations. The sine-waves resulting from a weighted sum of  
 the corresponding periodicities are overlaid in grey colour. Note that \pa\
 shows a much better match and a longer quasi-periodic coherence time scale than
 \pb. See Section \ref{qpdsect} and  Appendix \ref{apppcf} for details.}
\label{pcfplots}
\end{figure*}

\section{Quasi-periodic patterns}
\label{pcf}
\label{qpdsect}

In the previous section we elucidated the basic statistics of the null-burst
distributions in both pulsars. Both have bright phases whose lengths are
approximately normally (or possibly lognormally) distributed over lengths with a
similar number of pulses (Figs. \ref{BBB_lengh_gap_fig}a and
\ref{BBB_lengh_gap_fig}b), but their respective separations follow very different
statistics (Figs. \ref{BBB_lengh_gap_fig}b and \ref{BBB_lengh_gap_fig}d). In the case of
\pa\  the separations of the bright phases have a bimodal distribution,
suggesting that the pulsar must sometimes `skip' a burst (Fig.\ref{BBB_lengh_gap_fig}b), 
giving an exceptionally long off-phase. In \pb, the burst
separations cover a very wide range from 100 up to 1000 pulses 
(Fig. \ref{BBB_lengh_gap_fig}d). These features are apparent in Fig. \ref{spdisplay}, 
with the sequence of \pa\ including a very long off-phase, and that of \pb\
producing a quasi-periodic effect despite the varying separations.  

To probe deeper into the nature of ``quasi-periodicity" in our two pulsars, 
we require a suitable tool. Power spectra are a common device [e.g. \cite{hjr07}],
but in pulsars whose burst pulses form clusters such spectra are very much
dominated by red noise due to the observed jitter in cluster separation. In the
present context, a more useful procedure is to form a Pair Correlation Function
(PCF) for each pulsar. This is simply a histogram of all burst-to-burst
separations, whether successive or not, and can be utilized to find the
coherence time over which any quasi-periodicity is maintained. A formal
description of PCFs can be found in Appendix \ref{apppcf}. 

The PCF for \pa\ in Fig. \ref{pcfplots}a is formed from adding the
burst-to-burst separation histograms from all observed pulse sequences, 
where each sequence is approximately 2000 pulses long. 
The pulse clustering is very clear and reveals a dramatic
periodicity of about 170 pulses between each pulse of a cluster and the pulses
in other clusters. If all bright bursts were equal in length then the height of
the peaks would be equal for nearby separations and slowly decline for large
separations (see Appendix \ref{apppcf}), but we see a striking drop in the level
of these peaks for the two closest bursts to a given burst, followed by a strong
third peak. More distant peaks also have irregular levels, but a general
periodicity of about 170 pulses is maintained. In essence, it is the structure
revealed by this PCF which underlies the bimodal distribution of the burst phase
separations shown in Fig. \ref{BBB_lengh_gap_fig}b with its second peak at about 500
pulses. 

To understand our result we formed the Fourier spectrum (FFT) of the PCF, which
is shown in the inset diagram of Fig. \ref{pcfplots}a. This indicates two 
separate periodicities corresponding to approximately 170 pulses 
and 270 pulses, with the former dominating. The weighted sum of two sine-waves
with these periodicities is overlaid on the PCF and demonstrates a good match
with the PCF peaks. The 170 pulse and 270 pulse periodicities in the PCF are approximately the
third and the second harmonics respectively of $\approx 500$ pulse periodicity
reported by \cite{gjk12}. The peaks produced by the dominant periodicity of 170 pulses
are diminished for two successive peaks and then enhanced for the third by the
weaker but significant harmonically-related periodicity of 270 pulses. 
The weaker periodicity is not \emph{precisely} harmonically related to 170 and
thereby produces a progressive difference in the peak levels. What is very
remarkable is that this reproduces very closely the relative magnitude of the
peaks throughout the combined 2000 pulse separation of Fig. \ref{pcfplots}a.
This suggests an emission pattern which maintains coherence over at least 2000
pulses. However, we must caution that these results may or may not 
persist on timescales longer than our observations. 




The PCF for \pb\ (Fig. \ref{pcfplots}b) has its first peak at 
around 500 pulsar periods. This is more pronounced than that of \pa\, but much
broader in terms of pulses and it clearly corresponds to the peak found in the
bright phase separations (Fig. \ref{BBB_lengh_gap_fig}d). A second peak occurs at 1150
pulses, which is little late to be simply periodic with the first peak, and later
peaks show very little evidence of long-term coherence. We obtained the Fourier
spectra of the PCF,  which is shown in the inset diagram in Fig.
\ref{pcfplots}b, indicating three periodic features at 540, 595 and 490
pulses, with the first dominating. The weighted sum of the three sine-waves is overlaid
on the PCF but, in contrast to \pa, the generated wave loses coherence beyond 
1500 pulses as only the first two peaks are matched. Thus in \pb, three sine
waves are needed to yield just the two leading peaks of the PCF, in stark
contrast to \pa, where two sine waves were enough to match the entire
observation. This suggests that the decomposition into sine waves
has little physical significance in this pulsar.

It is indeed apparent in Figs. \ref{spdisplay} and \ref{BBB_lengh_gap_fig}d 
that the bursts of \pb\ appear with a wide variety of unpredictable separations. 
Thus the superficial impression of quasi-periodicity is only maintained by the fact that that the
separation of successive pulses is rarely less than 500 pulses, as is indicated
by the PCF. At two and three burst separations, there seems little evidence of
memory operating between bursts and even less of an underlying periodicity. 




\section{Pulse energy modulation in bright phases}
\label{BBB_patten}
\begin{figure}
\centering
 \subfigure[]{  
    \includegraphics[width=3.5in,height=3.2in,angle=0,bb=100 50 350 300]{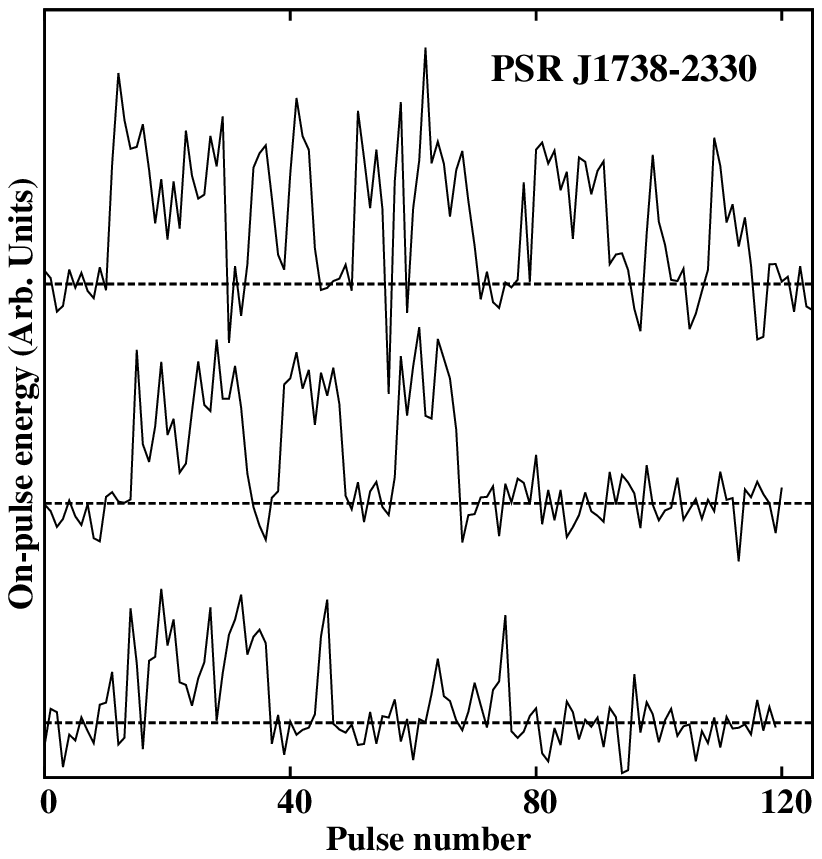}
    \label{j1738BBB}
  }
  \subfigure[]{  
    \includegraphics[width=3.2in,height=3.5in,angle=-90,bb=70 150 570 650]{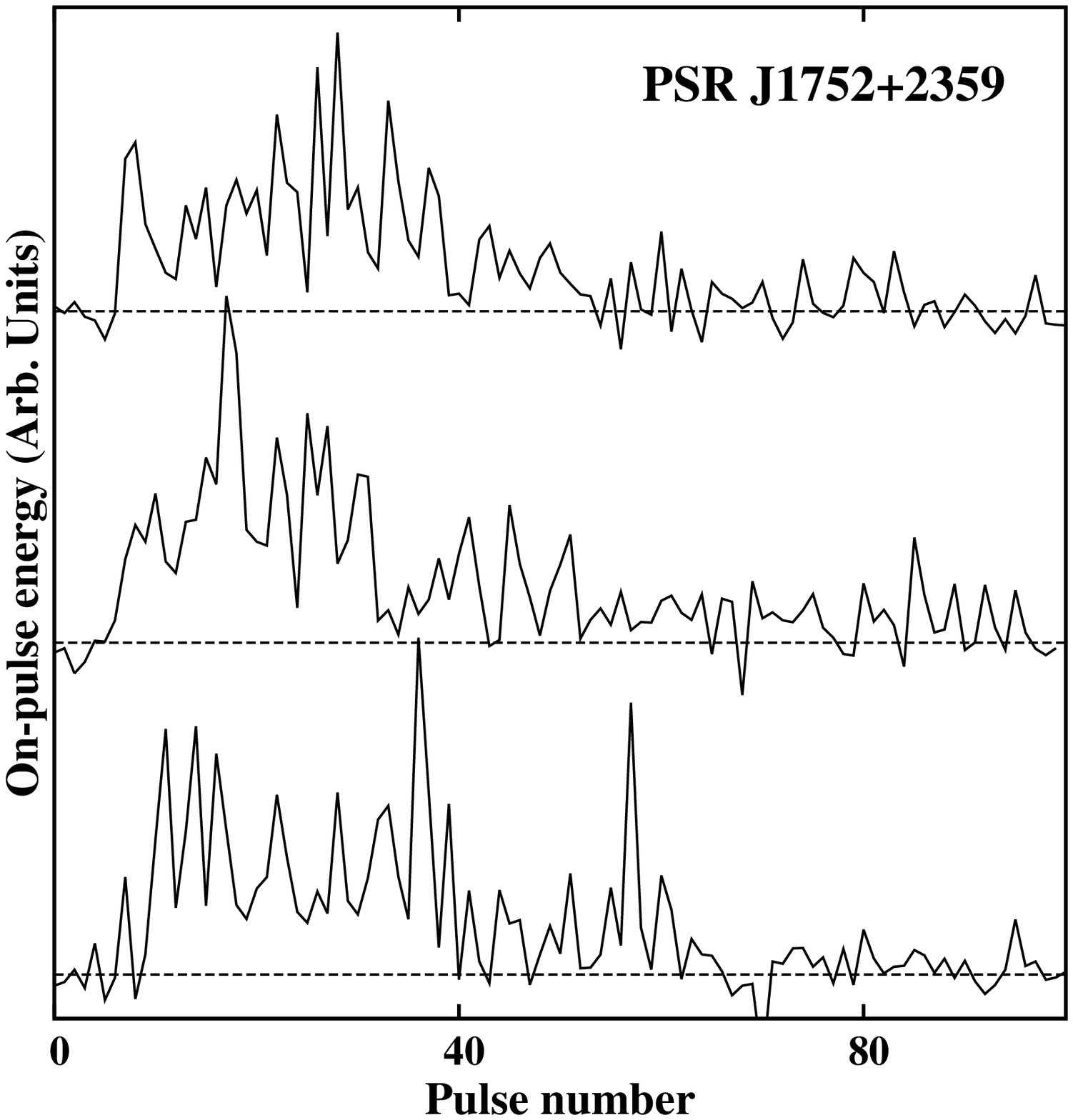}
    \label{j1752BBB}
  }
\caption{Examples of the on-pulse energy for three bright phases for PSRs (a) \pa\ (b) \pb\ 
 extracted from the GMRT observations. Both  pulsars show a gradual decay of 
 the on-pulse energy towards the end of the bright phase. 
 Note the short nulls within the bright phase for \pa. 
 By contrast, for \pb\ most pulses during a bright phase are not nulls.}
 \label{BBB_example}
\end{figure}
%

\begin{figure*}
  \centering
  \subfigure[]{
    \centering
    \includegraphics[width=2.2in,height=3.1in,angle=-90,bb=50 50 554 770]{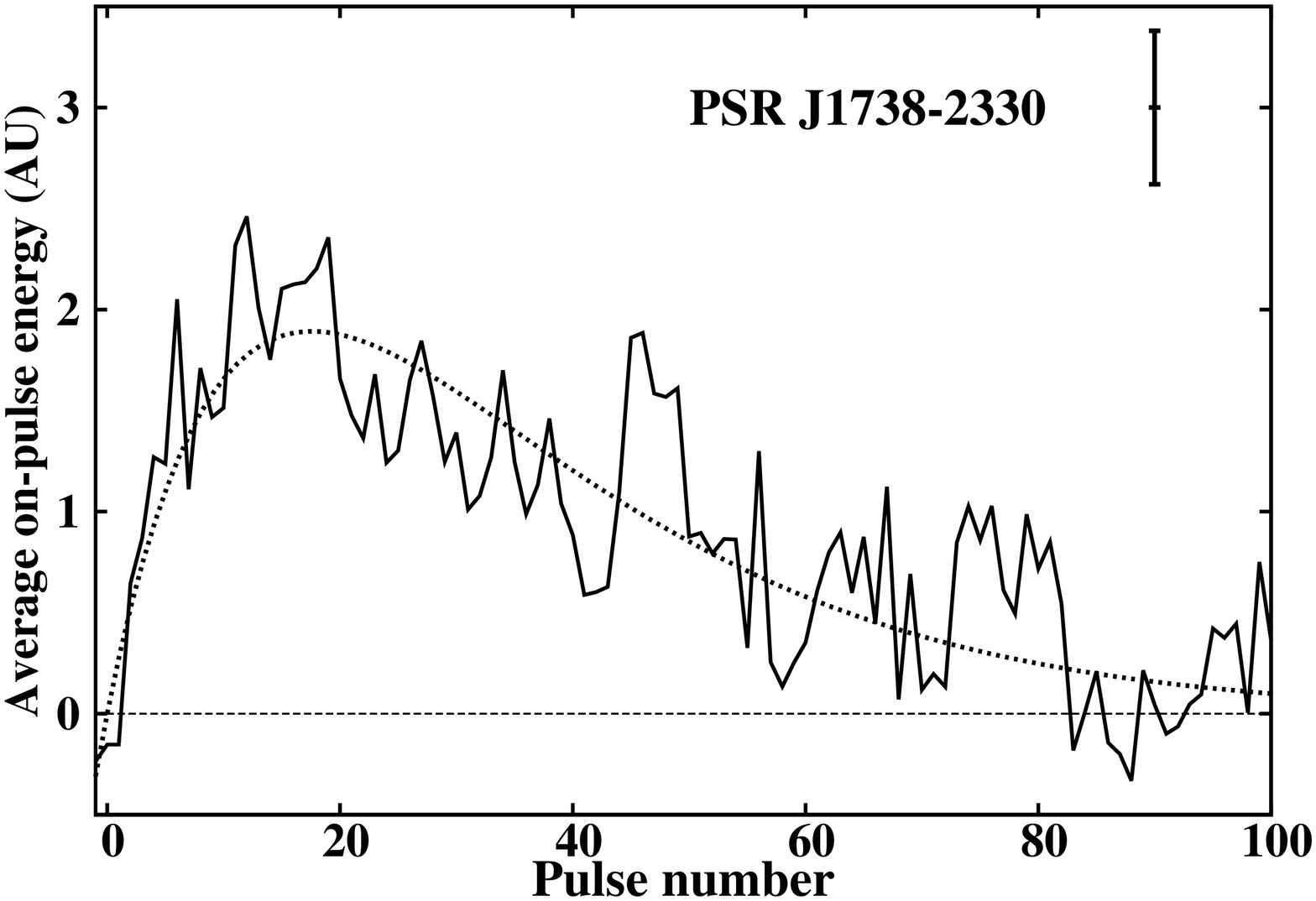}
    \label{j1738burst_fit} 
  }
  \centering
  \subfigure[]{ 
    \centering
    \includegraphics[width=2.2in,height=3.1in,angle=-90,bb=50 50 554 770]{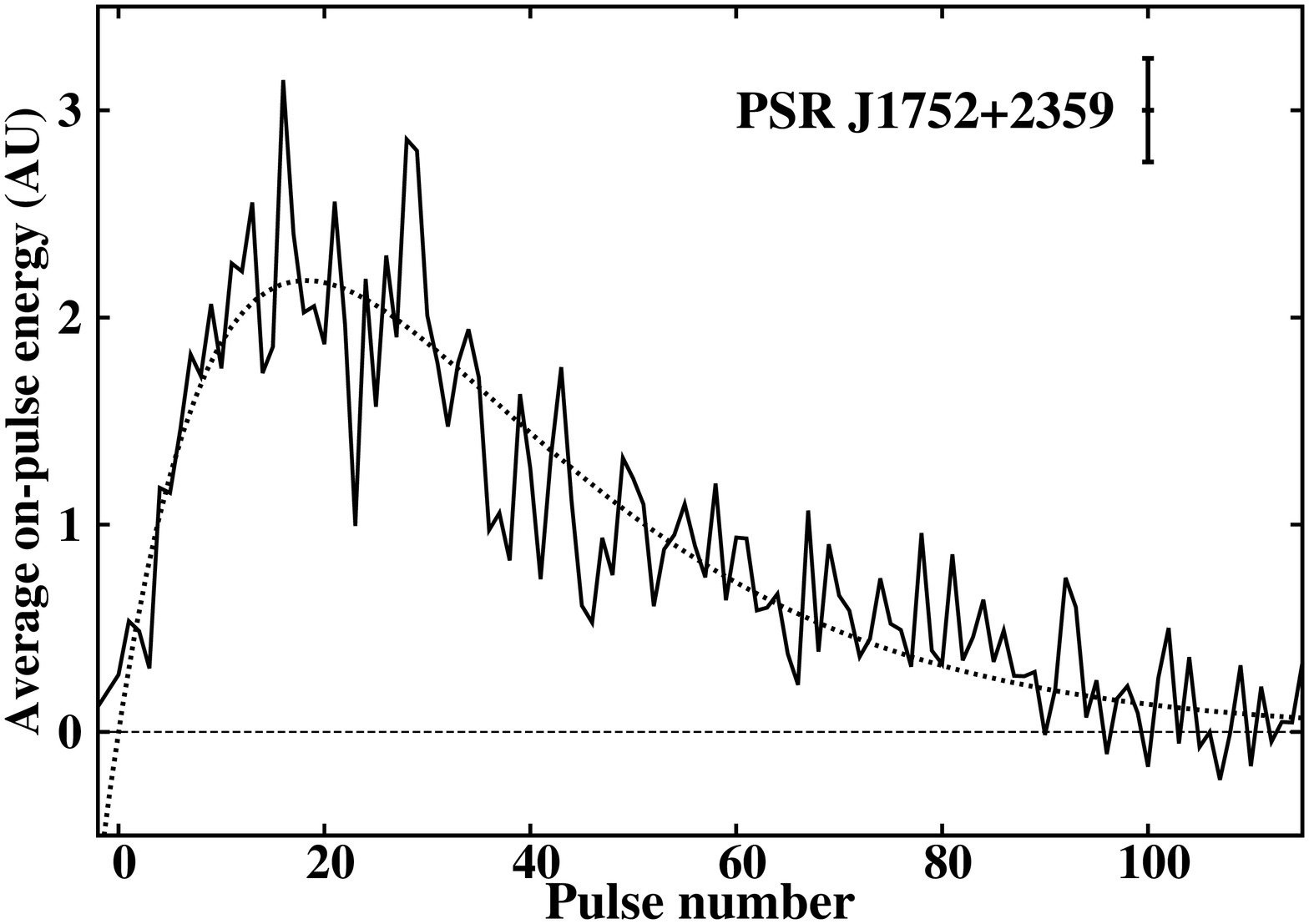}
    \label{j1752burst_fit_avg}
  }
 \caption{On-pulse energy for 120 pulses, averaged over 12 bright phases,
 for PSRs (a) \pa\ and (b) \pb\ in the GMRT observations, demonstrating 
 a similar decay in the bright phases of both pulsars.  
 The off-pulse root-mean-square deviations are shown in the top right corner for both
 pulsars. The dotted lines are fitted models given by equation \ref{fx1}, 
 with the reduced $\chi^2$ of around 1.2 and 1.9 for PSRs \pa\ and \pb, respectively.}
\label{burst_fit_avg}
\end{figure*}

\psra\ exhibits bright phase structures of various lengths, 
consisting of short burst bunches 
interspersed with short nulls, as was first reported by  
\cite{gjk12}. The onset of a bright phase is relatively sudden for \pa\ 
with a strong burst pulse, which is followed by a change in the 
emission throughout the bright phase duration. This change 
is manifested by either a reduction in the intensity of 
single pulses or by an increase in the number and/or 
length of short null states, as can be seen in the three 
examples displayed in Fig. \ref{BBB_example}a. At the end of every bright phase, 
the pulsed emission clearly goes below the detection threshold 
and produces long null phase or off-phase. We extracted 120 
pulses starting with the first identified burst pulse 
and averaged these over several bright phases to obtain 
averaged bright phase profile. 
A few bright phases in our observations were separated by less than 
120 pulses from the next consecutive bright phase, hence they 
were not included in this analysis. The on-pulse 
energy averaged over 12 bright phases is shown in 
Fig. \ref{burst_fit_avg}a, where a decline in 
the pulse intensity towards the end of the averaged bright phase 
is evident. 

The onset of the bright phase in \psrb\ is more gradual, 
spanning typically 5 to 10 pulses. The decline in the intensity 
from its peak is also more striking, as was 
also reported by \cite{lwf+04}. Fig. \ref{BBB_example}b shows three examples
of the decline in on-pulse energy during a bright phase in this pulsar. Note the absence of
convincing null pulses during the decline. The on-pulse energy for 120 pulses,
averaged from an equal number of bright 
phases (i.e. 12) as that for \pa, is shown in Fig. \ref{burst_fit_avg}b to illustrate the 
similarity of average bright phase on-pulse energy variations in the two pulsars. 

This variation in the bright phase on-pulse energy is well 
modelled by a functional 
form given in equation \ref{fx1} [also reported by \cite{lwf+04}]. 
\begin{equation}
f(x)~=~{\alpha}\cdot{x}\cdot{e^{-(x/\tau)}} 
\label{fx1}
\end{equation}
The average length of a bright phase can be derived from a least-squares 
fit to this functional form, as discussed in Appendix \ref{appa}. 
These were 86$\pm$4 pulses and 88$\pm$3 pulses for \pa\ and \pb, 
respectively and are consistent with the histograms of bright phase 
lengths discussed in Section \ref{qperiodsection} 
(Figs. \ref{BBB_lengh_gap_fig}a and \ref{BBB_lengh_gap_fig}c, which were obtained 
from the visual inspection). The length of the 
individual bright phase for \pb\ was obtained 
in a similar manner. Out of 123 observed 
bright phases in \pb, only 83, with higher signal-to-noise 
ratio (S/N) burst pulses, were fitted 
to obtain their lengths. We obtained the reduced  
$\chi^2$ in the range of around 0.6 to 2.2 for these fits. 
The average length of the bright phases from these measurements 
is 77$\pm$20 pulses. 

\begin{figure}
\includegraphics[width=2.7in,height=2.2in,angle=0,bb=30 50 390 302]{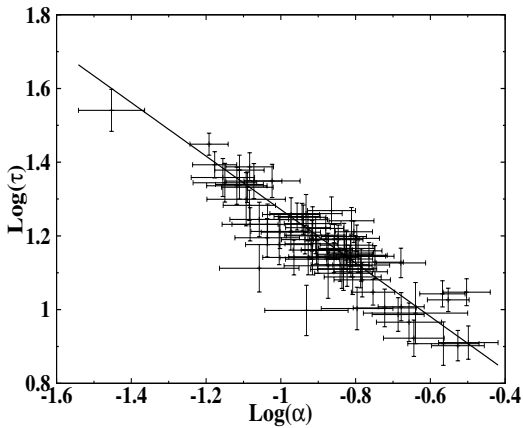}
  \caption{Relationship between bright phase parameters for \pb\ 
  obtained from the GMRT observations. The bright phase on-pulse energy decay
  timescale (i.e. $\tau$) as a function of peak of 
  the bright phase on-pulse energy (i.e. $\alpha$) on a log-log scales. 
  The slope was fitted after considering errors in both coordinates. 
  A strong anti-correlation is evident in this diagram.}
\label{bbb_beta_decay}
\end{figure}

In the case of \pb, we were able to investigate the nature of bright phases
further due to their sufficient number in our $\sim$ 8 hours GMRT observations  
as well as in high S/N 30 minute observations  with the Arecibo
telescope. The on-pulse intensities of around 83 observed bright phases 
were fitted to the model given by equation \ref{fx1} and 
their respective $\alpha$ and $\tau$  were obtained. 
The log-log plot in Fig. \ref{bbb_beta_decay} clearly displays a power law 
dependence of $\tau$ with $\alpha$. The fitted line in the Fig. \ref{bbb_beta_decay} 
gives the power law index of around $-$0.74$\pm$0.04  
incorporating the errors on both axes. The Kendall$\textquoteright$s tau 
rank order correlation between $\tau$ and 
$\alpha$ is around $-$0.67 with a very small 
probability ($<10^{-7}$) of random chance. Thus in \pb\ 
bright phases with large peak intensities decay faster.

We also investigated the relationship between the separation 
between consecutive bright phases with the  
parameters of bright phase preceding and succeeding the null phase/off-phase  
under consideration. We plotted the $\alpha$ and 
the $\tau$ of a bright phase as a function of the length 
of the off-phase preceding and succeeding it (not shown here). 
The lengths of the long off-phases were estimated 
as the number of pulses between the last and the first pulse 
of two consecutive bright phases (for example number of pulses between 
pulse numbers 2885 and 3421 in Fig. \ref{sp_en_oz_combined}).
We did not find any correlation as the parameters showed similar scatter 
for all lengths. Hence, bright phase parameters are independent of the 
length of the off-phase occurring before and after it. 

The strong anti-correlation between  $\alpha$ and  
$\tau$ suggests that the area under the on-pulse energy envelope 
for a bright phase for \pb\ is constant. 
This area and its error can be estimated for the assumed model 
(equation \ref{fx1}), as discussed in Appendix \ref{appa}.
These were calculated for the 83 observed bright phases and 
were the same, within errors, for all of them. This 
indeed confirms that the total intensity of bright phase is the same 
irrespective of its length or peak intensity and consequently, 
the total energy released during a bright phase is likely to be 
approximately constant.

\section{First and Last bright phase pulse}
\label{first_and_last_bbb_section}

\begin{figure}
 \centering
 \subfigure[]{
 \includegraphics[width=2.0 in,height=3.0 in,angle=-90,bb=50 50 554 770]{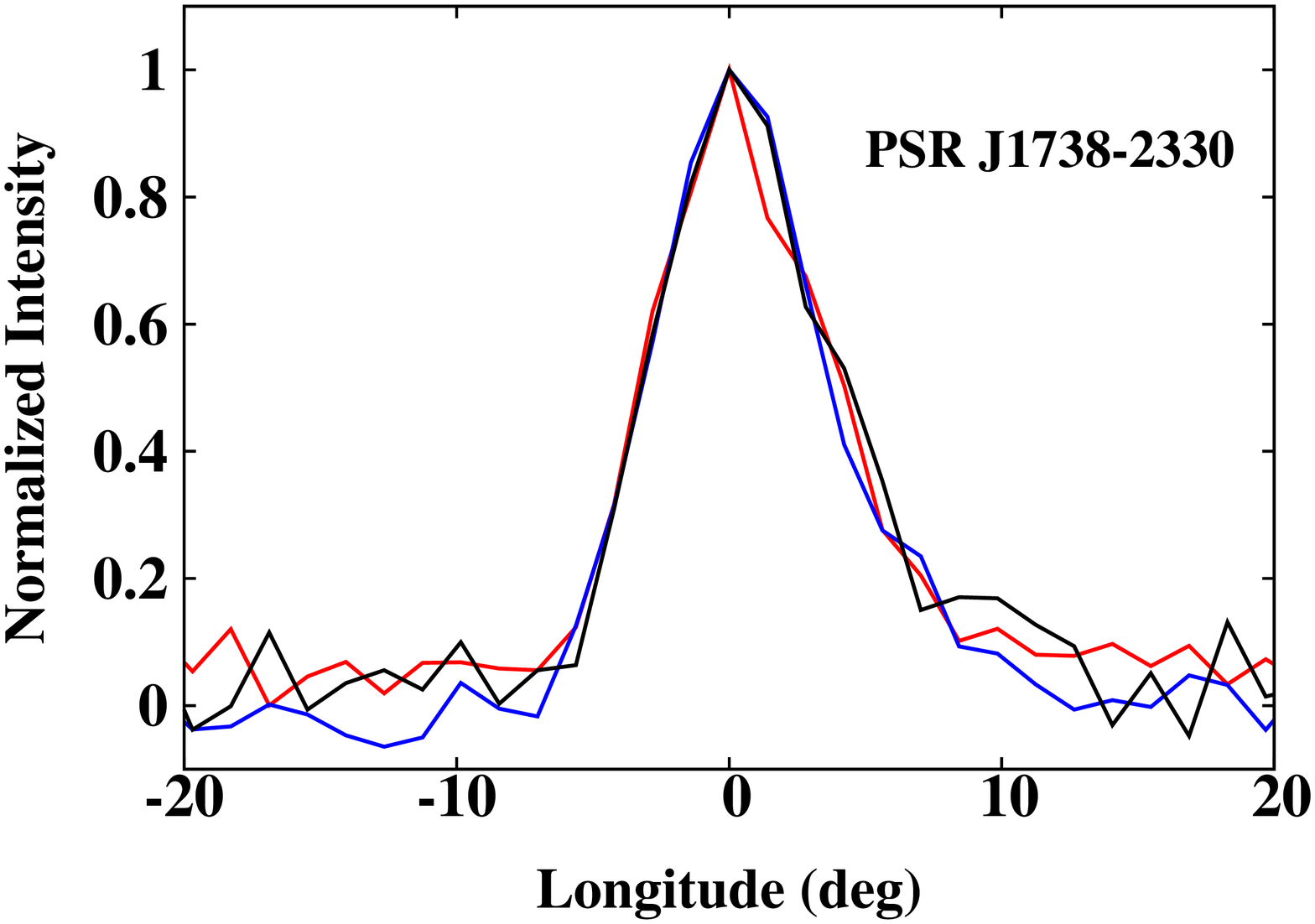}
 \label{j1738frst_last}
 }
 \subfigure[]{  
 \includegraphics[width=2.0 in,height=3.0 in,angle=-90,bb=50 50 554 770]{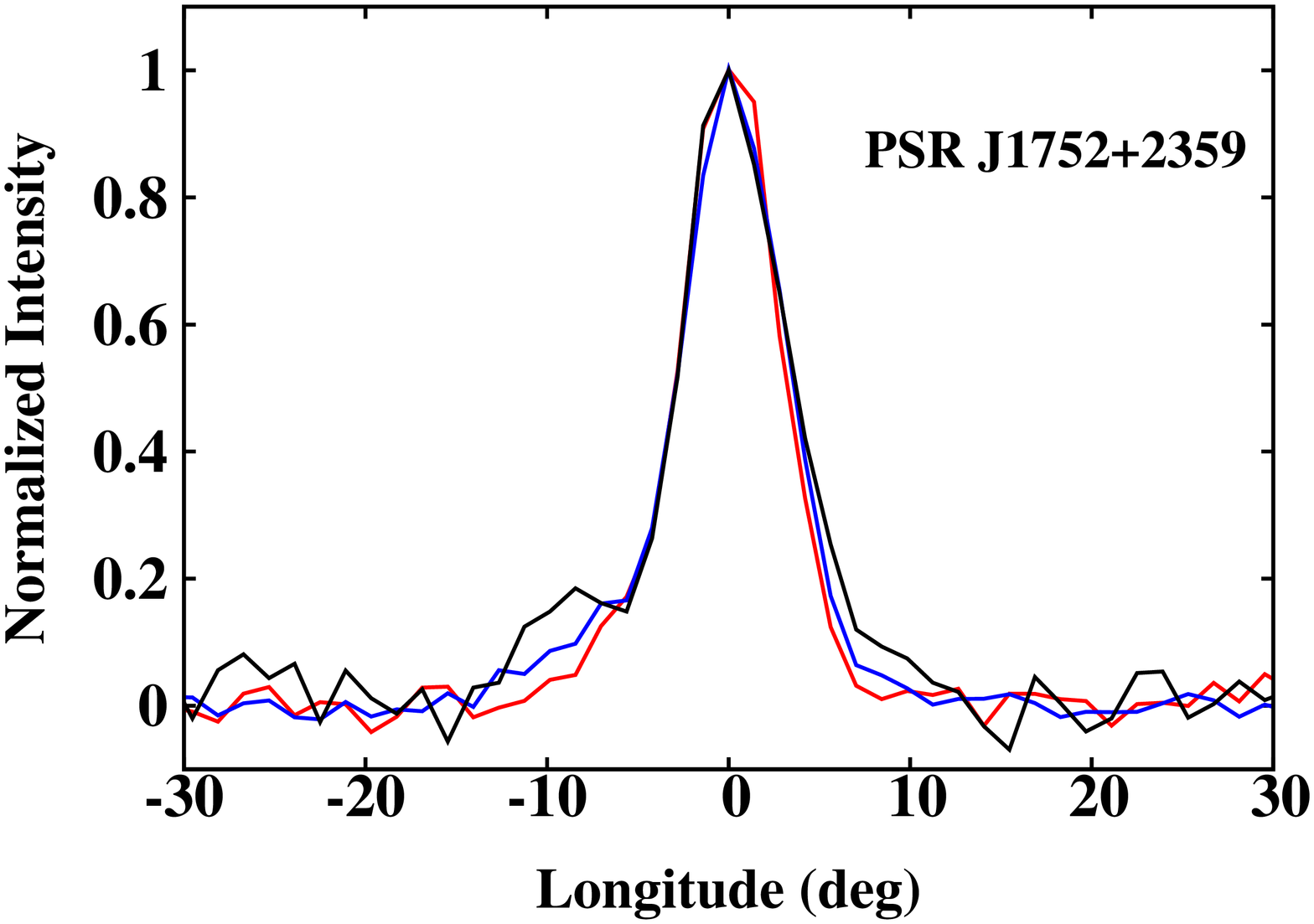}
 \label{j1752frst_last}
 }
 \caption{First-bright-phase-pulse profiles (red solid lines) and the last-bright-phase-pulse profiles 
 (black solid lines) obtained from the GMRT observations for PSRs (a) \pa\ and (b) \pb\ 
 plotted against the observed pulse longitude along with the the average pulse profiles (blue solid lines). 
 All profiles were normalised with their respective peak intensities for comparison.} 
 \label{first_last}
\end{figure}
Compared with \pb, \pa\ emits relatively strong individual pulses with
high S/N during the entire span of its bright phases (as can 
be seen from Fig. \ref{BBB_example}a). 
Although the intensity of the \pa's pulses during a bright phase shows 
a decline towards its end, the energy of the last pulses 
are sufficiently above the detection threshold to identify them 
clearly. The first and the last pulse of 18 bright phases were 
combined to form the first-bright-phase-pulse profile and 
the last-bright-phase-pulse profile. Fig. \ref{first_last}a 
shows these profiles for \pa\ along with its average pulse 
profile. All these profiles look similar. A KS-test comparison,  
carried out between the first-bright-phase-pulse profile and the average pulse 
profile,  indicates similar distributions with 
94\% probability. Similarly, a KS-test comparison 
between the last-bright-phase-pulse profile and the average pulse profile also suggests  
similar distributions with even higher probability of 99\%. 
These results suggest that \pa\ switches 
back into the bright phase mode from the off-phase 
(and vice-versa) without any significant change in the 
emission. However, these results should be treated 
with caution due to the small number of pulses available 
in forming first-bright-phase-pulse profile and last-bright-phase-pulse profile.
 
For \pb, only 114 out of 123 observed bright phases were used 
to obtain the first-bright-phase-pulse as the remaining were affected by RFI. 
The red solid line in Fig. \ref{first_last}b  
shows this profile. While the profiles look similar, a KS-test 
comparison between the first-bright-phase-pulse profile and the average pulse profile 
rejects the null hypothesis of similar distributions with 
99\% probability. The last pulse for most of the 
bright phases of \pb\ is very difficult to identify  
due to the decline in the pulse energy during this phase. 
Instead, a range of last pulses (10 to 20 pulses) were used for 
each bright phase to form the last-bright-phase-pulse profile. The length of each bright phase  
was determined with its associated error from the least-square-fit 
as explained in Appendix \ref{appa}. To obtain the last-bright-phase-pulse profile, we averaged 
a range of pulses within the error bars around the 
expected last pulse, obtained by adding the estimated length for every 
bright phase to the pulse number of its first pulse. 
The last-bright-phase-pulse profile averaged from 114 bright phases 
were compared with the average pulse profile 
using the KS-test, which rejected the null hypotheses 
of similar distributions with 99.9\% probability. 
Fig. \ref{first_last}b shows the last-bright-phase-pulse profile with  
a significant ($>$ 3 times off-pulse root-mean-square) 
component preceding the pulse and shoulder emission 
after the pulse. Neither of these features is present in its 
average profile. 

\section{Emission in the off-phase of \psrb}
\label{emission_in_null}
\begin{figure}
 \centering
 \includegraphics[width=3.5in,height=3.5in]{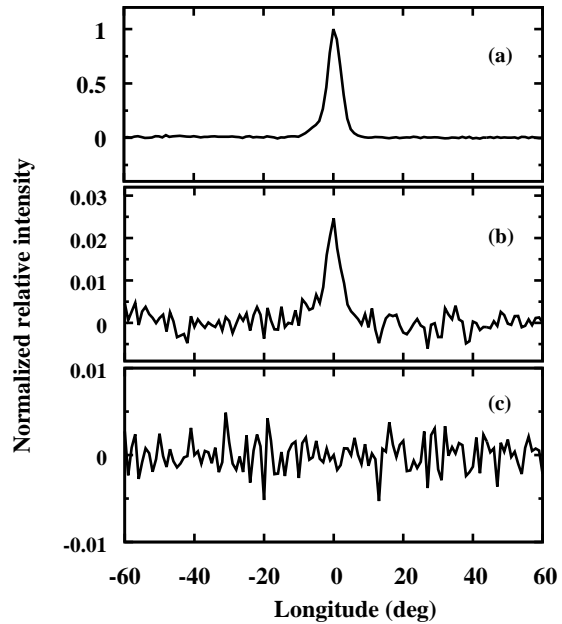}
 \caption{Plot of three profiles with the normalised intensities as a
 function of observed pulse longitude for \pb\ obtained from the GMRT observations.
 (a) The pulse profile obtained from all pulses in the bright phases (9000 pulses),
 (b) the off-phase pulse profile from all pulses between the bright phases
 (41500 pulses) and (c) the  `true' null pulse profile from the 38800 off-phase
 pulses which remain after removing the 2700 IBPs. All three profiles are on the same relative
 scale after normalising them with the peak intensity of the bright phase pulse
 profile.}
 \label{BBB_inter_null_profile}
\end{figure}
In Section \ref{quasiperiod} we noted that, unlike \pa, \pb\ exhibited
intermittent single pulse bursts during the off-phase. Weak emission
during an apparent null phase is not unknown \cite[]{elg+05} and may question
whether weak pulses are sometimes confused `true' nulls. 
We therefore investigated this by forming the off-phase pulse 
profile by averaging all pulses between the 
bright phases. Those parts of the observations  (about 15\%), affected by the RFI, 
were excluded from this analysis. All the pulses in identified bright phases, 
amounting to around 9000 pulses, were separated 
from the remaining single pulse observations . The profile obtained 
from these pulses (i.e. bright phase pulse profile) is shown 
in Fig. \ref{BBB_inter_null_profile}a. The remaining 
41,500 pulses, which occurred between the bright phases, were 
integrated to form the off-phase pulse profile. Surprisingly, the 
null-pulse profile showed weak emission  
with a significance of around 20 standard deviations ($\sigma$) (Fig. 
\ref{BBB_inter_null_profile}b). It is therefore important to clarify whether
this emission originates from bright but rare single pulses and/or from
underlying weak emission.

In many nulling pulsars, the burst pulses are strong enough to create a bimodal
intensity distribution and can then be separated by putting a threshold between 
the two peaks in the histogram \cite[]{gjk12,hjr07}. \psrb\ does not show such a bimodal
distribution due to the presence of many weak energy pulses (Fig. \ref{hist_both_fig}b). 
Hence, the use of any threshold will lead to a wrong identification 
of weaker burst pulses as nulls and result in 
a weak emission profile during the off-phase, as seen in Fig. \ref{BBB_inter_null_profile}b.
This could be due to (a) weak emission throughout the off
phase, (b) emission from weak burst pulses in the 
diminishing tail of a bright phase, or (c) emission from a few wrongly
identified individual burst pulses during the off-phase. 

\begin{figure*}
 \centering
 \includegraphics[width=5.5in,height=3.2in,angle=0,bb=50 50 410 302]{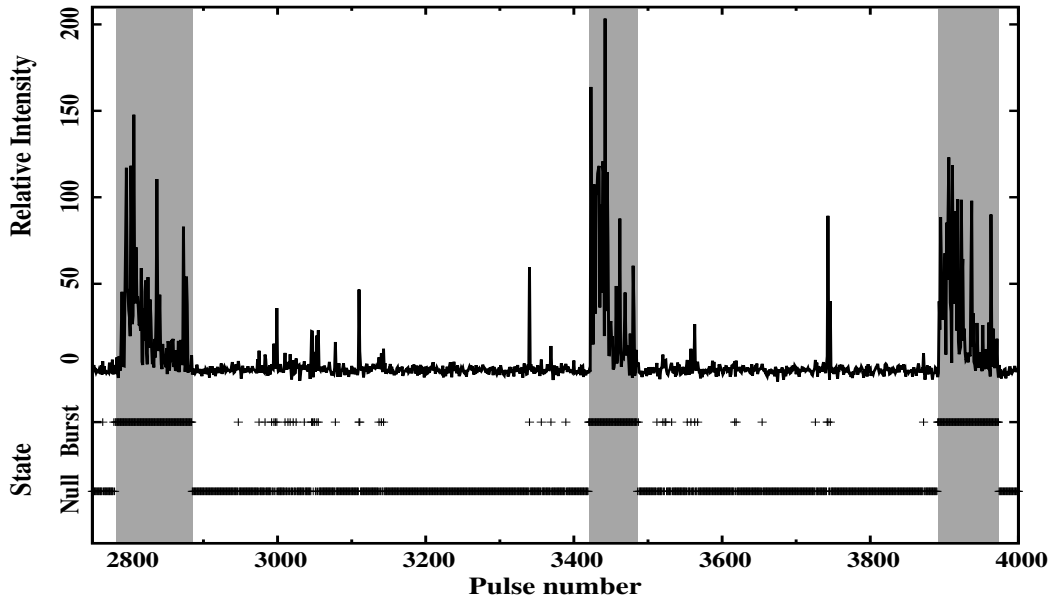}
\caption{Plot of on-pulse energy variation of around 1200 pulses of \pb\ observed
at 327-MHz with the Arecibo telescope is shown in the top panel. The three
identified bright phases are shown inside the grey shaded regions for clarity. 
Every individual pulse was classified either as a null or as a burst pulse. The
bottom panel shows the null or burst state for the corresponding pulse 
consistent with a random distribution of the IBPs.}
\label{sp_en_oz_combined}
\end{figure*}

To distinguish between the above possibilities, we 
used higher S/N observations  for this pulsar obtained from the Arecibo telescope. 
Fig. \ref{sp_en_oz_combined} shows a section of the
observations  from the Arecibo observation. The pulse energy plot clearly 
shows three bright phases with about 60 isolated single 
burst pulses occurring during the off-phase between them. 
This strongly supports our earlier suspicion that weak emission,   
seen during the off-phase in this pulsar, is due to these 
isolated burst pulses (i.e. pulses outside the bright phases or IBPs). 

To investigate this further with GMRT observations, all pulses remaining after 
separating the bright phases were arranged in ascending order of their on-pulse
energy. A threshold was moved from the high energy end towards the low energy end till 
the pulses below the threshold did not 
show an average profile with a significant (above 3$\sigma$) 
component.  All the pulses below this threshold 
were tagged as null pulses. They were again visually checked 
for wrongly identified nulls due to  
presence of low level RFI. The average profile, obtained from 
all the pulses tagged as null after this procedure, 
is shown in Fig. \ref{BBB_inter_null_profile}c, 
indicating that tagged pulses are all nulled pulses,  
ruling out the possibility of weak emission throughout the null. 

This method of identifying the IBPs was also used on 
the Arecibo observation. The null or the burst state for each individual 
pulse was identified and is shown in Fig. \ref{sp_en_oz_combined}   
in the bottom panel. It can be seen from this figure 
that IBPs are not localised near either the start or the end 
of a bright phase, but are distributed randomly inside the long off
phase. In fact, these pulses may not be confined to off-phases and may occur at
random through all phases since within a burst phase the their numbers would be
small and not detectable. To the best of our knowledge, 
the presence of such pulses has not been reported in a pulsar before.

\begin{figure}
 \centering
 \includegraphics[width=3 in,height=2.2 in,angle=0,bb=50 50 410 302]{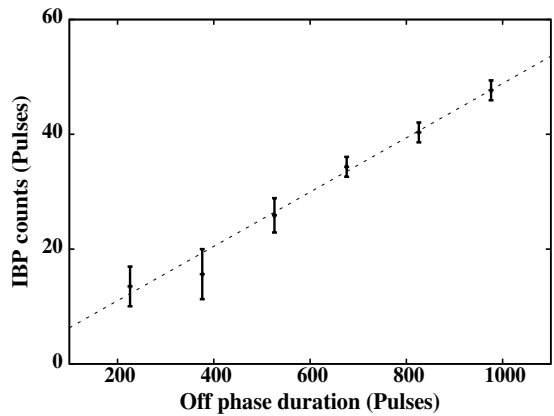}
 \caption{Plot of inter-burst pulse (IBP) counts as a function of length of
 off-phase  between two consecutive bright phases as measured from the GMRT 
 observations of \pb. A linear relation is evident between the IBP count 
 and the corresponding off-phase length, indicating that the IBP rate is 
 independent of duration of off-phase.}
 \label{burst_rate_hist}
\end{figure}

A random occurrence of IBPs implies that the rate of IBPs 
should remain constant.  The Arecibo observation were short and hence 
statistical analysis of the IBP rate was not possible. 
Using GMRT observations we identified around 2700 such pulses, 
using the variable threshold method described above. 
As discussed in Section \ref{BBB_patten}, the length of every 
bright phase  was derived after fits to equation \ref{fx1}. 
An acceptable fit could not be obtained for a few bright phases 
due to either low pulse energy or due the presence of strong RFI. 
We only considered those off-phases bounded at both ends by 
bright phases, with acceptable fits. 
A few such off-phases were also affected by RFI and hence 
not included. The number of identified IBPs were 
counted for each off-phase and the IBP rate estimated using about 53 such off
phases of various lengths. Fig. \ref{burst_rate_hist} shows 
a histogram of the number of IBPs for a given off-phase 
between two consecutive bright phases. The error bars on them were obtained 
from Poisson statistics. Fig. \ref{burst_rate_hist} clearly 
demonstrates that the number of  IBPs 
are linearly correlated with the corresponding off-phase lengths. 
Hence, the IBP rate is independent of the length of a given null phase and could
remain fixed throughout the entire emission of the pulsar. 
%
%
\section{Comparison of polarization profiles}
\label{Arecibo}
\begin{figure}
 \centering
 \includegraphics[width=3.5 in,height=2.6 in,angle=0,bb=50 50 410 302]{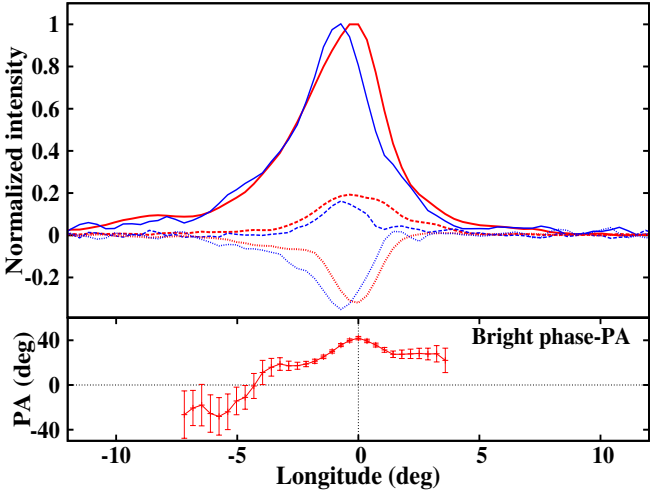}
 \caption{Plot of average polarization profiles
  from a half-hour observation of \pb\ at 327-MHz using the Arecibo
  telescope. The total intensity profile for bright phase pulses (red solid line) and IBPs
  (blue solid line) are both normalised to their peak intensities. The linear
  polarization profiles for bright phase pulses (red dashed line) and IBPs (blue dashed line) 
  are normalized by the peak intensity from their respective total intensity
  profiles. The circular polarization profiles for bright phase pulses (red dotted
  line) and IBPs (blue dotted line) are also normalized in a similar manner. 
  The bottom panel shows the derived position angle (red solid line) as a
  function of pulse longitude for bright phase pulses.  A clear offset is 
  evident in the total intensity and the circular polarization profiles between 
  the profiles for bright phase pulses and the IBPs.}
\label{pol_comp}
\end{figure}
We analysed the full polar observations from the Arecibo telescope, observed 
with 25-MHz bandwidth at 327-MHz.  
The bright phase pulses and the IBPs were separated from the observed  
single pulse observations and separate polarization profiles were obtained 
for both. 
Fig. \ref{pol_comp} shows the total intensity profiles along with 
the linear and the circular polarization profiles for the bright 
phase pulses and the IBPs. The position angle  swing was also
measured for longitude bins where the observed linear polarization 
was more than 3 times the off-pulse rms in the linear polarization profile. 
%

Fig. \ref{pol_comp} shows striking differences between the intensity and
polarization profiles of the bright phase pulses and the IBPs. The total intensity profile 
of the IBPs is clearly shifted to earlier phase with respect to that of
bright phase. A KS-test comparison between these two profiles rejected the null hypothesis of similar 
distributions with 99.9\% probability. The average intensity of the IBPs 
is around 5 to 7 times weaker than the average bright phase pulses. 
A Gaussian function was fitted on both the profiles to estimate the 
position of their peaks.  The offset between the peaks in the total 
intensity profiles for bright phase pulses and the IBPs was estimated  
to be around 0.6$\pm$0.1$^{\circ}$.  The linear polarization 
profile for the bright phase pulses is wider than that for IBPs and 
shows strong linear polarization of around 20\% as compared 
to about 16\% for IBPs. The shift in the position 
of peak intensity between these two profiles 
is not very significant. The circular polarization 
profile of the IBPs is offset from that of the bright phase pulses
by  0.54$\pm$0.07$^{\circ}$. However, in contrast to the overall reduction 
in the pulse energy during the IBPs, the IBP circular polarization 
fraction shows a small increase compared to that for the bright 
phase pulses (circular polarization of 35\% and 32\%, respectively) 
as is also evident from Fig. \ref{pol_comp}.

\section{Summary of results}
\label{conclusion}

\begin{table*}
{\small
 \caption{A list of comparisons and differences between the two pulsars.}
 \centering 
 \begin{center}
 \begin{tabular}[ht]{|l|c|c|}
 \hline 
     &   \psra\       & \psrb\ 	   \\
 \hline
 \hline
 Null Fraction (NF)  &   85.1(2.3)\%  & $<$ 89\%     \\ 
 Quasi-periodicities & 170 and 270 pulses  & 540 (490,595) pulses \\
 Coherence of quasi-periodicities & $>$ 2000 pulses (11 peaks) & $\sim$ 1000 pulses (2 peaks) \\ 
 Variation in on-pulse energy during bright phase  &  Exponential decline with flickering nulls  & Steady exponential decline  \\
 Length of bright phase & 86 $\pm$ 4  pulses         & 88 $\pm$ 3 pulses \\
 Separation between bright phases & $\sim$ 170 and 500 pulses    & $\sim$ 570 pulses \\
 Inter-burst pulses (IBPs)       & No (?)                         & Yes; with a fixed random rate \\
 First and last bright phase pulses  & Similar to each other and to average profile & Distinct from each other and from average profile \\ 
 \hline
 \end{tabular}
 \end{center} 
 \label{table_comp}
 \hfill{}
 }
\end{table*}
As we have stressed from the start, the superficial similarity in the emission
patterns of  \pa\ and \pb\ hides many differences. \psra\ is the younger
pulsar but with a stronger magnetic field, which has enabled it to spin down to
a period five times longer than the older, but weaker, \psrb\ (see Table \ref{paratable}).
Nevertheless, both have arrived at a point where their nulling fractions are
greater than 80\% and their bursts of emission are of similar length 
(measured in pulses) separated by long null sequences which give the overall appearance of
being quasi-periodic (Fig. \ref{spdisplay}, Table \ref{table_comp}).

However, detailed examination of the patterns of emission bursts reveals
significant differences. Firstly, the burst separations of the younger pulsar
\pa\ turn out to be underpinned by periodic behaviour. A histogram of the
burst pulse separations (PCF in Fig. \ref{pcfplots}) could be modelled by the
superposition of two near-harmonically related sine-waves and matched to the
observations for at least 2000 pulses (i.e 11 burst phases). This is strong
evidence of long-term memory in this pulsar. However, although the coherence of
the sine waves is not lost, individual burst phases often fail to materialise
except in some vestigial form. This can be seen as the effect of the stronger
sine periodicity sometimes countered by the effect of the weaker near-harmonic
(2:3) periodicity.

By contrast, in \pb\ there is no evidence of long-term periodicity. This
pulsar requires a complex wave superposition to reproduce just the first two
burst phase intervals in its pulse-separation histogram  (Fig. \ref{pcfplots}b). 
Beyond about 1000 pulses, coherence is lost rapidly and there is no memory of any
periodicity. The impression of quasi-periodicity is maintained since the burst
separations usually range between 300 and 600 pulses (Fig. \ref{BBB_lengh_gap_fig}d).

Additionally, \pb\ has a striking feature not present in the younger pulsar --
and hitherto not reported in any pulsar. In the long null intervals between the
bursts of \pb\  the nulls are interrupted at random by burst pulses which are
mostly single and relatively weak in intensity (IBPs). These pulses have a
profile significantly shifted with respect to the main profile and different
polarisation properties (Fig. \ref{pol_comp}), hinting at a different physical
origin. They may well occur continuously through both burst and off-phases,
providing a diffuse background to the pulsar's more structured emission of
bright and off-phases. 

The bright phases of the pulsars, although similar in duration when measured in
pulses, have different substructures. In \pa, the bursts start suddenly and
as they progress, they are increasingly punctuated by nulls until the off-phase begins
(Fig. \ref{BBB_example}a). Thus the onset and the end of the bright phases are clearly
marked, and profiles obtained by integrating first and last burst pulses are
effectively identical to the pulsar's overall profile (Fig. \ref{first_last}a).
In \pb, the bursts take longer to reach their peak emission, followed by an
exponential decline in intensity. The bright phase fades with weak emission and
it is often difficult to pinpoint its true end (Fig. \ref{BBB_example}b). There is
evidence that an additional small leading component appears in the emission
profile as the phase closes (Fig. \ref{first_last}b).

In both pulsars the progression of the intensity of their burst phases can, when
averaged, be fitted by the same functional form (equation \ref{fx1}, see Fig.
\ref{burst_fit_avg}). In the case of \pb, we find that individual bright
phases with a higher peak intensity tend to be shorter in length with a
significant anti-correlation ($-$0.7) between these two parameters 
(Fig. \ref{bbb_beta_decay}), implying that the output in energy integrated over a
bright phase is constant, at least  at the observing frequency. Furthermore, the
bright phase parameters are independent of the length of the off-phase before
and after the bright phase.

\section{Discussion and Conclusions}
\label{discussion}
It has long been known that no strong correlations exist 
between NF and other pulsar parameters such as period or 
period derivative \cite[]{rit76,ran86,big92a,wmj07}. 
This is also true here since, despite their similar NFs, 
the positions of PSRs \pa\ and \pb\ in the $P-\dot{P}$ diagram 
could hardly be further apart. Furthermore, in this same 
diagram both are far from the so-called 
``death line" -- contradicting a simple view that 
pulsars die through progressive increase in NF \cite[]{rit76}. 
However, what high NF pulsars do seem to have in 
common is that their individual null pulses do not appear at 
random with a single fixed probability. 
In fact, their few non-null pulses have a tendency to cluster 
in what we have called ``bright phases", even when these phases are separated by 
hours or even days \cite[]{klo+06,lem+12}.  

The simplest way of generating clusters of bright emission 
is to assume a two-state probability model, such that the 
probability of a null during a bright phase is fixed but 
lower than that during an off-phase \cite[]{gjk12,cor13}. 
This represents an elementary Markov (Poisson point) process and 
results in separate exponential distributions for the null and 
burst lengths, so that short nulls/bursts are very common 
and long nulls/bursts much less likely. This model has been 
plausibly applied to the null and burst statistics of several 
pulsars with low to medium NFs [see Figure 3 of \cite{gjk12} and 
Figure 1 of \cite{cor13}]. In all these models burst/null clustering occurs and their 
power spectra show broad red features, but they do not generate quasi-periodic features.

However, in the case of the two pulsars described in this paper we have become 
convinced that neither of them has null/burst distributions which are exponential 
in form. This is most evident in the burst distribution of \pb\ (Fig. \ref{nbhist_both}c), 
which appears to be bimodal, and when the IBPs are disregarded (as in Fig. \ref{BBB_lengh_gap_fig}c) 
a simple Gaussian-like distribution becomes clear. The same is true of the pulsar's 
apparently exponential null distribution in  Fig. \ref{nbhist_both}d, which, 
on neglect of the IBPs, is transformed into a single hump in Fig. \ref{BBB_lengh_gap_fig}d. 
In \pa, the burst/null distributions of Figs. \ref{nbhist_both}a and \ref{nbhist_both}b 
are superficially exponential, but we know them to partly consist of short bursts and 
nulls which arise exclusively during the bright phases, 
again meaning that the distribution is bimodal. When these short nulls/bursts 
are disregarded, as in Figs. \ref{BBB_lengh_gap_fig}a and \ref{BBB_lengh_gap_fig}b, 
a single-hump distribution is evident (in fact a double 
hump in the case of Fig. \ref{BBB_lengh_gap_fig}b for reasons given in Section 3). 
This line of reasoning is similar to that of Kloumann \& Rankin (2010) 
in their study of the high NF pulsar B1944+17. They suggest that the shortest 
nulls of that pulsar are ``pseudo-nulls", which in fact integrate to a weak profile, 
leaving a hump-like distribution for the remaining nulls.

We are therefore forced to abandon the simple (Poisson point) assumption of a 
separate but fixed null probability for each of the two phases. Instead, we 
see that the probability of the length of time the two pulsars spend in their 
bright and off-phases is dependent on its respective Gaussian-like distribution. 
It is this which gives the pulse sequences their quasi-periodic character, as 
typical bright phase and off-phase alternate with lengths scattered around 
the means of their distributions. 

Within this picture, the burst phases will not occur in precisely periodic sequences, 
but can be expected to gradually lose coherence as their separation varies randomly about a mean.
This can be seen very clearly in the PCF of \pb\ (Fig. \ref{pcfplots}b), where coherence is 
lost after only two peaks.  In \pa, the typical burst phase separations 
are less and their distribution narrower (Fig. \ref{BBB_lengh_gap_fig}b), so we might 
expect coherence to persist for more peaks, as is indeed the case (Fig. \ref{pcfplots}a).  However,
the degree of coherence in this pulsar's PCF is 
exceptionally high and we have been able to reproduce it well by combining 
just two near-harmonic underlying periodicities. This hints at the presence, 
at least in this pulsar, of forcing periodicities with suitable phase shifts \cite[]{cor13}. 
How this is achieved physically is a matter of speculation. For example, these periodicities may arise from 
an external body \cite[]{csh08} or neutron star oscillations \cite[]{cr04a} 
or near-chaotic switches in the magnetosphere's non-linear system \cite[]{tim10}.


In both pulsars the bright phases themselves show clear evolution and therefore 
have some kind of internal memory (Figs. 6 and \ref{burst_fit_avg}). 
Both diminish in intensity and \pa\ has an 
increasing number of null pulses as the burst proceeds. 
A decline of energy during a burst has been 
noted in several other pulsars. Recently, \cite{lem+12} 
have found that  PSR J1502$-$5653 has long nulls interspersed with weakening 
bursts (their Figure 2) whose peak drifts as the burst develops. 
A similar phenomenon appears to occur in PSR J1819+1305 \cite[]{rw08} 
and \cite{bgg10} have also reported a gradual fall in pulse intensity before 
the onset of null states for PSR B0818$-$41. In other studies, \cite{ysw+12} 
report intermittent long off-phases in PSR B0823+26 and their Figure 1 clearly 
shows a decline in intensity before null onset. In all cases it seems that the 
off-phases do not come out of the blue, so that the burst phase may represent 
a reset or relaxation of the magnetospheric conditions. It is clear that a 
change-of-state occurs when the pulsars move to the off-phase, quite possibly 
involving a global magnetospheric change \cite[]{ckf99,con05,tim10}.

The appearance of random isolated pulses (IBPs) during the off-phase of \pb, 
exhibiting a different integrated profile to the burst profile, has not been 
reported in other long-null pulsars. Its random nature is reminiscent of RRATs 
and the bright single pulses which appear in the weak mode of PSR B0826$-$34 
\cite[]{elg+05} or the RRAT phase of PSR J0941$-$39 \cite[]{bb10}. 
We cannot know if this emission represents an additional property of the stable but 
intermittent off-phase magnetospheric state or whether it is a permanent background 
phenomenon such as accretion \cite[]{w79}, which has a separate physical origin.

The differences in the statistics and the structure of the bright phases of 
PSRs \pa\ and \pb\ do not necessarily imply that the two pulsars produce nulls 
in a fundamentally different way. It is possible that through having a 
stronger surface magnetic field and a wider light cylinder \pa\ is 
somehow able to maintain near-periodic coherence for longer than \pb\ 
(considerably longer if measured in clock time rather than pulse numbers), 
and possible differences in the unknown inclinations of the pulsars' dipole 
axes to their rotation axes may play a role \cite[]{csh08}. 
Our results suggest that pursuing in detail the ``quasi-periodic"  behaviour of 
any pulsar may well yield valuable physical clues to the nature of its 
magnetosphere and environment.

An earlier study of pulsars with \emph{low} NFs \cite[]{gjk12} found 
that the NF percentage was not predictive of a pulsar's subpulse behaviour. 
Our detailed study of PSRs \pa\ and \pb\ has shown that pulsars with very 
similar, but \emph{high}, NFs can also have subtle but important differences 
in emission.

\section{Acknowledgements}
We would like to thank the staff of the GMRT and NCRA for providing
valuable support in carrying out this project.
The authors also thank Joanna M. Rankin for sharing 
the calibrated polarization observations for \psrb\ with us. 
VG would like to thank Dipanjan Mitra for valuable discussion 
regarding the Arecibo data analysis and nulling in general. 
The authors are grateful to the referee for a critical review of the paper and suggesting 
several improvements in the manuscript. BCJ thanks A. Timokhin 
for very useful discussions on different magnetospheric  states. 
GW thanks the University of Sussex for a Visiting Research Fellowship.

\bibliographystyle{mn2e}
\bibliography{psrrefs,modpsrrefs,mybib,modrefs}

\appendix
\section{No Giant pulses from PSR J1752+2359}
\label{gps}
\begin{figure}
 \centering
\psfig{figure=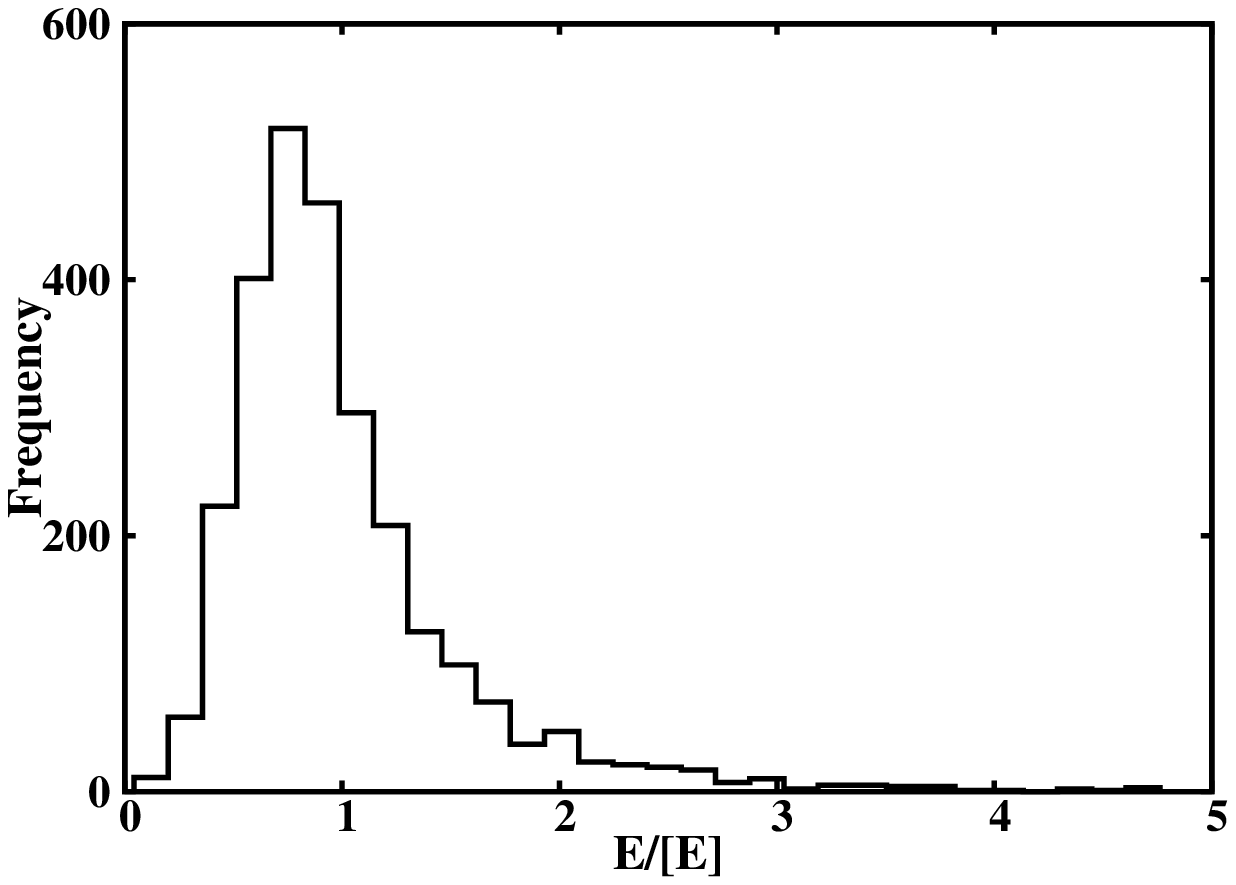,height=2.5in,width=3.5in,bb=50 50 410 302}
\caption{The 0n-pulse energy histogram of pulses with S/N $\geq$ 5 
with the on-pulse energies scaled as explained in the text.}
\label{ESPs}
\end{figure}

PSR J1752+2359 is one of the few pulsars which shows 
giant pulses (GPs) at 111-MHz. 
The energy of these GPs was reported to be 200 times 
the energy of the mean pulse \cite[]{EK05}.  
Our observations show only a gradual distribution 
up to 20 times the mean pulse energy (Fig. \ref{hist_both_fig}a).
We separated all single pulses with  S/N greater than 5 
and estimated the mean pulse energy from these well defined burst
pulses. The distribution of on-pulse energies of all burst pulses, 
scaled by this mean pulse energy is shown in 
Fig. \ref{ESPs}, indicating that these are distributed  
only up to 5 times the mean pulse energy, which is significantly lower 
than that reported by \cite{EK05} at 111-MHz.  
In the earlier studies of GPs \cite[]{sr68,kt00,J04} it was reported 
that (a) GPs have significantly smaller pulse widths compared to 
average pulses and (b) GPs tend  to occur mostly at the edge of 
the average pulse profile. We separated the strongest pulses to 
compare their widths with that of the average pulses and no 
significant difference was found. While the profile for IBPs 
is slightly narrower and shifted towards the leading edge, their 
intensities are 5 to 7 times weaker than the average bright phase pulses. 
If the IBPs have different spectral behaviour compared to bright phase pulses, 
then it is likely that they may give rise to GPs at 111-MHz. 
However, this needs to be tested with simultaneous observations 
at both frequencies. Hence, it can be concluded from our observations   
that J1752+2359 does not show GPs at 325-MHz. 
\section{The Pair Correlation Function}
\label{apppcf}
The Pair Correlation Function ({\itshape PCF}) is a probability density
function 
(also known as the radial distribution function or pair separation function) for
the clustering 
of certain objects or events in space and/or time coordinates \cite[]{plu85}  
and is useful for measuring the degree of packing. We have 
used a one-dimensional PCF, which 
identifies the clustering of events (the bursts pulses of a bright 
phase in our case), in the time series data. A brief  
description is provided here as this 
seems to be the first time such a technique is applied 
the clustering of burst pulses in pulsar astronomy. 

The PCF for a series M pulses 
with N burst pulses  can be derived as follows.  
The pulse index of these burst pulses are 
\begin{equation}
 p_i~ or ~ p_j~=~p_1,~p_2,~...,~p_N.
\end{equation}
Then,  PCF is defined as 
\begin{equation}
 g(p)~=~G\cdot\sum\limits_{i=1}^{N}\sum\limits_{j\neq{i}}^{N}\langle{\delta(p~-~\arrowvert{p_j-p_i}\arrowvert)}\rangle.
 \label{pcfeq}
\end{equation}
where G is a scaling parameter and $\delta$ is the Kronecker delta function.
A normalized binning of PCF, g(p), provides the probability  
of occurrence of certain separations between burst pulses. If a 
pulsar exhibits bunching of burst pulses and periodic  
occurrence of these bunches, the PCF shows prominent peaks 
around repeatedly occurring separations and their 
harmonics. A simple way to detect such periodicities is to 
obtain the Fourier spectra  of the PCF.

\begin{figure}
 \includegraphics[width=3.5 in,height=2.5 in,angle=0]{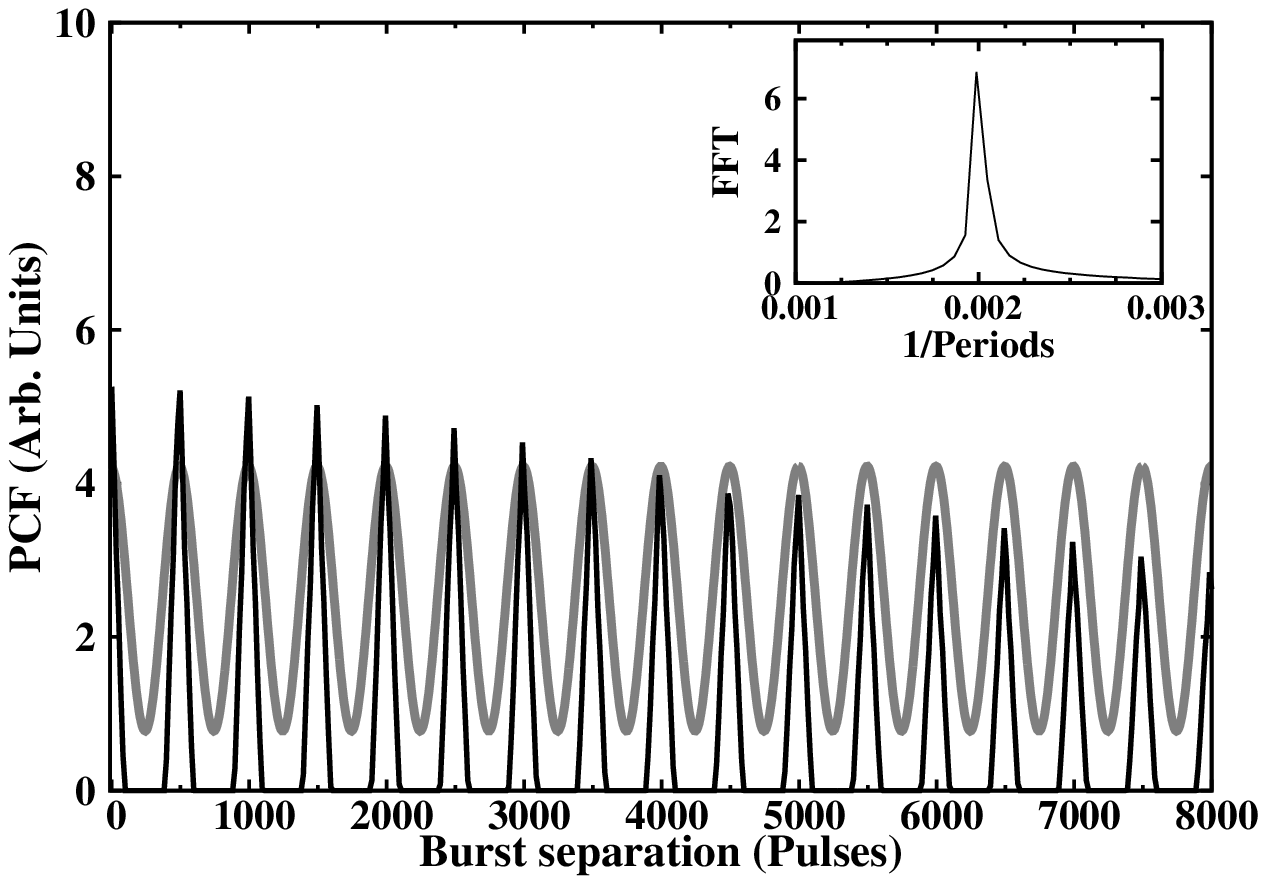}
  \caption{PCF of a pulsar whose nulls form clusters of 100 pulses separated
by 400 periods. It is a histogram of the separations between the simulated
burst 
  pulses in units of pulse periods. Note the gradual 
  reduction in amplitude for large separations due to the finite 
  length of the simulated data. The inset figure shows the 
  Fourier spectra of the PCF. A sine-wave with a 500 pulse periodicity 
  is overlaid along with the PCF in grey colour for clarity.} 
  \label{Apndx_pcf}
\end{figure}

Fig. \ref{Apndx_pcf} shows an example of  the PCF obtained 
for 8000 pulses of a pulsar nulling with a precise periodicity. The pulsar was 
simulated 
by  repeated occurrence of 100 burst pulses 
separated by 400 null pulses, so the NF was 80 \%. 
The periodicity is clearly visible both in the PCF 
and in the Fourier spectra in the inset figure. The peaks in PCF are at 500
pulses, broadened by the 100 pulse spread.

The PCF measures not only the periodicity of the clustering but also how long
its coherence persists. If coherence in the bunching 
is lost, the PCF would not show peaks beyond a particular length. 
This makes it superior to a simple Fourier analysis of 
the pulse energy modulation since a PCF emphasises short-lived periodic
features 
as well as providing information regarding the coherence 
length. An additional advantage over conventional Fourier analysis 
is that observations from different sessions can be combined. The maximum coherence 
length measurable in this case will come from the session with 
the longest pulse sequence among all the  observing sessions. 
Hence, the PCF is a useful technique to scrutinize 
periodic pulse energy fluctuations. 

\section{Modelling of on-pulse energy variations in bright phases}
\label{Burst_pattern_section}
\label{appa}
Our analysis of \pb\ clearly shows that on-pulse energy for 
most of the individual bright phases follows the model given 
by equation \ref{fx}. \psra\ also has a similar average on-pulse energy 
variation for its bright phases. 
\begin{equation}
f(x)~=~{\alpha}\cdot{x}\cdot{e^{-(x/\tau)}} 
\label{fx}
\end{equation}
Here, $\alpha$ is a scaling parameter and $\tau$ is the decay 
time-scale. 
The values for $\alpha$ and $\tau$ were obtained by a least-square-fit 
of function \ref{fx} to the on-pulse energy in a bright 
phase with errors on each energy measurement given by 
off pulse rms.  The length of a given bright phase 
 was defined as the difference between 
the two points where $f(x)$ crosses  of 10\% of the $f(x)_{max}$.
It can be shown that $x_{max}$ is given by $\tau$. So the points 
where the $f(x)$ attains 10\% of the peak value (i.e. $x=x_{10}$) 
are given by 
\begin{equation}
{f(x)}\arrowvert{_{x = {x_{10}}}}~ = ~ 
{\alpha}\cdot{x_{10}}\cdot{e^{-(x_{10}/\tau)}} ~ =
~{0.1}\cdot\frac{\alpha\cdot\tau}{e}. 
\label{f10}
\end{equation}
%
%
$f(x)$ attains these values on both sides of the peak position 
and these points can be determined by solution of equation \ref{f10}, 
obtained using numerical methods and the difference between these 
two points was defined as the length of a given bright phase 
(i.e. $L$ =  $x_{10h}$ $-$ $x_{10l}$). It can be seen 
from equation \ref{f10} that values of $x_{10h}$ and $x_{10l}$ do 
not depend upon $\alpha$. To quantify the dependence of the 
bright phase length on $\tau$, we solved equation \ref{f10}  
for a range of $\tau$ values. We found that the 
length of a given bright phase is related to $\tau$ as 
\begin{equation}
 L ~ \approx ~ 4.9\times\tau. 
\end{equation}
%
%
The error on L is given by 
\begin{equation}
\bigtriangleup{L} ~ \approx ~ 4.9\times\bigtriangleup\tau. 
 \label{f10err}
\end{equation}
The area for the on-pulse energy variation in a bright phase is given by 
\begin{equation}
 A_{BBB}~=~\int_0^\infty
\mathrm{\alpha}\cdot{x}\cdot{e^{-(x/\tau)}}\,\mathrm{d}x 
 =~\alpha{\cdot}\tau^2
\end{equation}
and the error in the obtained area is given by 
\begin{equation}
\bigtriangleup{A_{BBB}}~=~\bigtriangleup{\alpha}\cdot{\tau^2}~+~2{\alpha}\cdot{
\tau}\cdot{\bigtriangleup{\tau}}.
\end{equation}
\end{document}